\documentclass[english]{sbc2025}%

\usepackage[misc,geometry]{ifsym} 

\usepackage{aas_macros}
\usepackage[bottom]{footmisc}
\usepackage{tabularray}
\usepackage{afterpage}
\usepackage{url}
\usepackage{pifont}
\usepackage{balance}

\setcitestyle{square}

\newenvironment{cds}{\subsection*{Citation Diversity Statement}\small}{}

\definecolor{engtitle}{rgb}{0.5,0.5,0.5}
\definecolor{orcidlogo}{rgb}{0.37,0.48,0.13}
\definecolor{unilogo}{rgb}{0.16, 0.26, 0.58}
\definecolor{maillogo}{rgb}{0.58, 0.16, 0.26}
\definecolor{darkblue}{rgb}{0.0,0.0,0.0}
\hypersetup{colorlinks,breaklinks,
            linkcolor=darkblue,urlcolor=darkblue,
            anchorcolor=darkblue,citecolor=darkblue}

\copyrightstatement{This work is licensed under a Creative Commons Attribution 4.0 International License}
\jyear{2026}

\category{Research Paper}

\title[Deceptive Patterns as a Sociotechnical Phenomenon]{Deceptive Patterns as a Sociotechnical Phenomenon: Review, Catalog, and Discussion}%

\author[Baroni et al. 2025]{
\affil{\textbf{Luiz Adolpho Baroni}~\orcidlink{0000-0003-3913-5976}~~[~{Federal University of Paraná}~|\href{mailto:labaroni@inf.ufpr.br}{{\textit{labaroni@inf.ufpr.br}}}~]}

\affil{\textbf{Alisson Andrey Puska}~\orcidlink{0000-0001-9965-3770}~~[~{Federal University of Paraná}~|\href{mailto:alisson.puska@inf.ufpr.br}{{\textit{alisson.puska@inf.ufpr.br}}}~]}

\affil{\textbf{Roberto Pereira}~\orcidlink{0000-0003-3052-3016}~~[~{Federal University of Paraná}~|~\href{mailto:rpereira@inf.ufpr.br}{{\textit{rpereira@inf.ufpr.br}}}~]}

}

\begin{document}

\begin{frontmatter}

    \maketitle




    \begin{abstract} 
    \textbf{Abstract:}
    \noindent \textit{Background:} Deceptive patterns are interface design strategies aimed at misleading users or favoring specific interests, compromising user experiences and ethical privacy principles. These patterns involve exploit different issues of the interaction between humans and technology, revealing gaps in the scientific literature regarding their understanding and working mechanisms. \textit{Purpose:} This research characterizes deceptive patterns as a sociotechnical phenomenon, integrating human and technical dimensions while offering a self-explainable interactive catalog to raise awareness among users and designers. \textit{Methods:} To characterize deceptive patterns as a sociotechnical phenomenon, a systematic literature review in Computer Science was conducted, and the Semiotic Framework was applied to analyze and organize the sociotechnical aspects of identified patterns in an integrated manner. An exploratory evaluation of the interactive catalog was conducted, combining heuristic evaluation and focus group. \textit{Results:} The research identified multiple deceptive patterns in scientific literature. Analysis revealed that existing studies frequently addressed specific aspects of patterns, such as typologies or effects, and do not focus on sociotechnical issues. The exploratory evaluation suggested that the self-explainable catalog is useful and easy to use, with potential to inform users and promote awareness regarding the deceptive patterns existence and functioning. \textit{Conclusion:} The study contributes to an informed understanding of deceptive patterns, highlighting the need for an approach that integrates human and technical dimensions. The self-explainable catalog is a promising tool to inform and promote awareness about the topic. By exposing the risks and mechanisms of these patterns, the research seeks to promote awareness among users and designers towards ethical interface design practices.
    \end{abstract}


\end{frontmatter}

\section{Introduction}
\label{sec:intro}

Deceptive and manipulative design patterns \citep{BRIGNULL2023} can be understood as design decisions intentionally applied to technical artifacts, whether digital or physical, to benefit a third party at the expense of those who use these artifacts \citep{CONTI2010,GRAY2018,MATHUR2021a}.

Deceptive patterns, such as cookie notifications that hide opt-out options or multi-step cancellation processes, like the 'Roach Motel,' manipulate users across digital and physical interfaces, causing frustration, privacy breaches, and financial losses. Although the literature recognizes these wide variety of issues, it struggles to fully grasp their nature due to a fragmented approach where studies often isolate social effects, like user manipulation, from technical mechanisms, such as interface design choices, rather than examining how they intertwine.

Deceptive patterns can emerge in any interface type, negatively impacting user behavior by prioritizing third-party interests and potentially leading to undesirable situations, negative emotions, or financial losses. Fragmented approaches hinders a comprehensive understanding, which is essential to address deceptive patterns as a sociotechnical phenomenon. A solely technical perspective fails to uncover the intentions behind deceptive interfaces, while a purely social analysis overlooks the mechanisms-from software to physical constraints, such as benches with bars-that operationalize them. While existing literature acknowledges challenges in addressing deceptive patterns, research has predominantly focused on specific aspects of these patterns, either focusing on the social or technical aspects.

Developing a holistic understanding becomes challenging without comprehensively characterizing deceptive patterns from a perspective that integrates social and technical considerations. Such an approach helps to comprehend deceptive patterns as technical artifacts (implemented through hardware and software) that actively shape and constrain human behavior, ultimately impacting social dynamics. 

Results from a systematic mapping presented in this paper revealed that existing research predominantly concentrated on narrow aspects of deceptive interfaces, such as cataloging and categorizing deceptive pattern types, presenting user perspectives, or discussing ethical and societal implications. These studies often ignore key questions: Who created or funded these patterns? How do companies exploit transmission mediums to manipulate users? How do these patterns leverage even the user's physical environment?

Existing studies \citep{GRAY2021b, GRAY2021, WESTIN2021, MATHUR2021a} cross disciplinary boundaries, mainly incorporating insights from Law and Psychology. However, these studies do not consider technical issues or implementation details that characterize these patterns. Moreover, these studies do not rely on a structured sociotechnical framework, potentially overlooking relevant social (user manipulation, societal impacts) and technical (interface design, transmission medium characteristics, physical world implications) dimensions.

The method adopted for our research is grounded in a conceptual-constructive approach, inspired by the framework proposed by \cite{oulasvirta2016hci} for understanding HCI research as a problem-solving process. Using analytical and constructive investigations, this research aims to achieve an understanding of deceptive patterns through a sociotechnical lens and to develop a technical artifact to present and organize these patterns in an easy way.

Therefore, the general objective of this research is: \textbf{To characterize deceptive patterns as a sociotechnical phenomenon in Computing}. Specific objectives:
\begin{enumerate}
    \item Analyze and characterize a deceptive pattern from a sociotechnical perspective, in order to fully comprehend these patterns as technical artifacts that directly influence the social world;
    \item Map the literature on deceptive patterns in computing, to understand how these patterns are being studied on the literature;
    \item Analyze the sociotechnical aspects considered and neglected in the literature, to grasp if and what points of these patterns researches are focusing on;
    \item Develop a catalog of the mapped deceptive patterns, to create an educational, self-explained, tool that can be used for developers and users to better understand, and protect themselves from, these deceptive patterns.
\end{enumerate}


While several studies have explored the social or technical sides of deceptive interfaces, few have integrated these dimensions under a unified sociotechnical lens. This work aims to bridge that gap.

\section{Background and Theoretical Foundations}
\label{sec:background}

In 1987, Kent and Cunningham adapted the concept of design patterns \citep{ALEXANDER1979} to computer programming \citep{BECK1987}. These design patterns were created to help software design practitioners solve already known problems in the software community: patterns comprising well-defined and documented practices or solutions for known software design problems. A well-known example of a user interface design pattern is \textit{"Good Defaults"}\footnotemark{}. When users need to fill out a form on a website or application, the values they use are often the same as other users, so the interface comes with preset values that make form completion easier. \footnotetext{Available at: http://ui-patterns.com/patterns/GoodDefaults Accessed: February 22, 2026.}

\subsection{Anti-patterns vs. Deceptive Patterns}

Besides software patterns, there are also anti-patterns (\textit{anti-patterns}), which are known poor practices or solutions that attempt to solve problems but end up being inefficient or counterproductive. The term anti-pattern was coined by Andrew Koenig in 1995 in his book "\textit{Patterns and anti-patterns}" \citep{KOENIG1998}. Hanlon's Razor\footnotemark{} applies to anti-patterns: "never attribute to malice that which is adequately explained by stupidity." \footnotetext{Available at: https://www.britannica.com/topic/Hanlons-razor Accessed: Feb 20, 2025.} A commonly used anti-pattern example is \textit{"Ambiguous link labeling"}\footnotemark{} -- e.g., when a site intends to redirect users using a link labeled "Click here!". Many users do not understand how the action should be performed or where the link will redirect. Anti-patterns are poor design decisions due to the designer/developer's ignorance or inexperience. \footnotetext{Available at: http://ui-patterns.com/blog/User-Interface-AntiPatterns Accessed: February 22, 2026.}

In 2010, Harry Brignull coined the term \textit{"dark patterns"}, which are "tricks used in websites and apps that make you do things that you didn't mean to, like buying or signing up for something" \citep{BRIGNULL2023}. However, the author abandoned the term "dark" and is now using \textit{deceptive and manipulative} to avoid language that contain negative associations or reinforce harmful stereotypes -- the ACM (Association for Computing Machinery) itself recognizes this term as inappropriate \footnote{Available at: https://www.acm.org/diversity-inclusion/words-matter Accessed: Feb 23, 2026}. Brignull also created a website, www.deceptive.design/, listing 12 types of deceptive patterns he found throughout his career as a user experience designer, including Trick Questions, Roach Motel, Privacy Zuckering, and others. Throughout this work, the term deceptive patterns is used as an umbrella term to encompass malicious interfaces and manipulative design patterns.


A key difference between anti-patterns and deceptive patterns lies in \textit{intentionality}: while an anti-pattern is flawed due to designers' lack of experience or knowledge, deceptive patterns are intentionally designed to deceive and exploit users or favor specific parties' interests (software owner or designer) at the users' expense. Another difference is \textit{complexity}: while many deceptive patterns show greater complexity in construction and manipulation, using various techniques and elaborate interfaces, anti-patterns tend to be less complex, products of their authors' lack of knowledge.

\begin{figure}[h]
\centering
\includegraphics[width=0.40\textwidth]{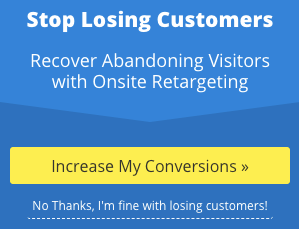}
\caption{Deceptive pattern known as \textit{Confirmshaming}.}
\label{fig:confirms}
\end{figure}

\begin{figure}[h]
\centering
\includegraphics[width=0.4\textwidth]{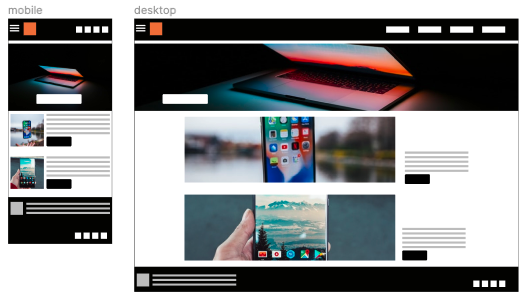}
\caption{Mobile Design On Big Screens: Anti-pattern where the developer/designer used a mobile \textit{layout} for websites without considering large screens.}
\label{fig:antip}
\end{figure}

Figure \ref{fig:confirms} shows a deceptive pattern known as Confirmshaming, which attempts to dissuade users from clicking on advertisements using emotional appeal. Figure \ref{fig:antip} shows a more recent anti-pattern example, Mobile Design On Big Screens, where developers create mobile-oriented layouts that stretch to fill desktop screens without considering the excess white space.\footnotemark

\footnotetext{Available at: https://keymedium.com/top-10-web-design-mistakes-in-2021-with-jakob-nielsen/. Accessed: February 22, 2026.}


Deceptive patterns are constructed to influence user behavior \citep{ACQUISTI2017} to benefit stakeholders other than the users themselves \citep{UTZ2019}. Studies show users are manipulated even when aware of manipulation \citep{BONGARDBLANCHY2021}. Academic research has explored the ethical implications of deceptive patterns \citep{GRAY2018,LACEY2019,CHIVUKULA2019a,DGERONIMO2020} as they can cause discomfort, frustration, anger, and other emotions \citep{MAIER2020,GRAY2021}. Other works have shown that these patterns manipulate behavior by exploiting cognitive biases \citep{MATHUR2019}, like the "Framing effect" \citep{TVERSKY1981}, where humans reach different conclusions from identical information based on presentation. For example, information can be presented to reduce user autonomy \citep{UTZ2019}.

\begin{figure}[h]
\centering
\includegraphics[width=0.4\textwidth]{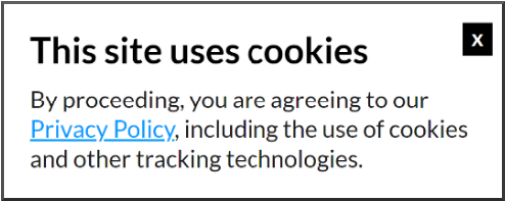}
\caption{Deceptive pattern of type \textit{Does not count}\citep{SOE2020}.}
\label{fig:cookie}
\end{figure}

Figure \ref{fig:cookie} shows a deceptive pattern example: a cookie notice stating, "By proceeding you agree to our Privacy Policy, including the use of cookies and other tracking technologies."\footnotemark{} This specific pattern is called "\textit{Does not count}" as it is an invalid way to obtain user consent.

\begin{figure}[h]
\centering
\includegraphics[width=0.4\textwidth]{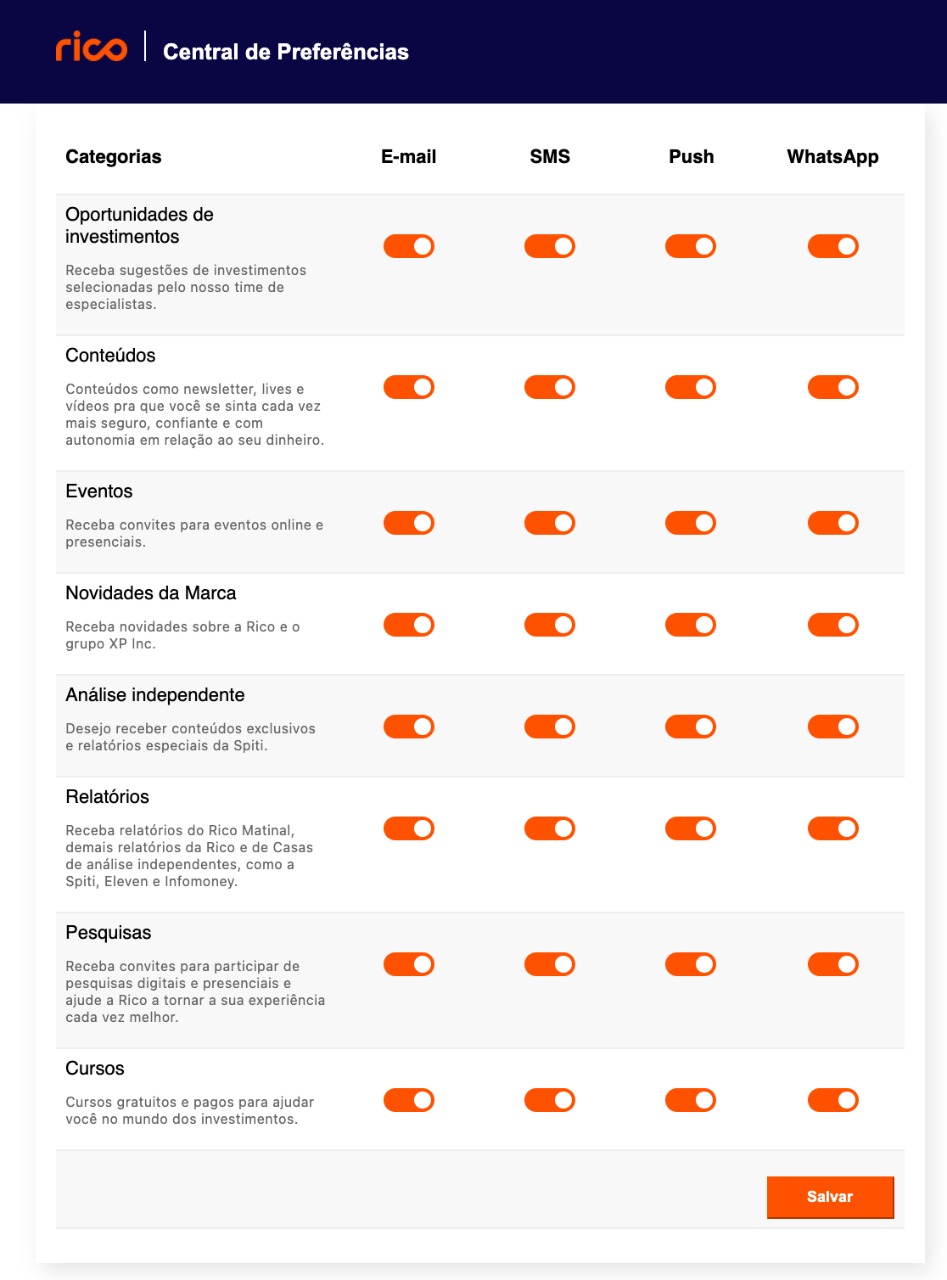}
\caption{Example of Obstruction.}
\label{fig:obstrucao}
\end{figure}

Other example of deceptive patterns is the Obstruction pattern presented in Figure \ref{fig:obstrucao}: the notification control interface requires clicking 32 buttons to turn off all notifications.

\begin{figure}[h]
\centering
\includegraphics[width=0.4\textwidth]{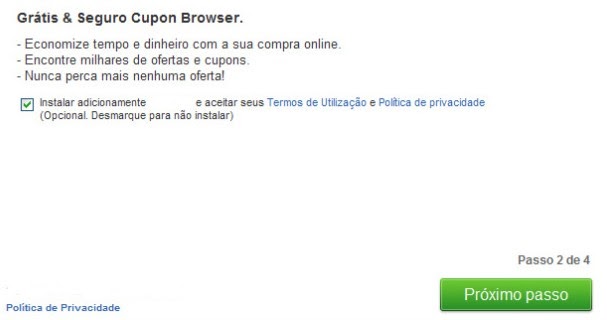}
\caption{Example of \textit{Sneaking}.}
\label{fig:sneaking}
\end{figure}

Figure \ref{fig:sneaking} demonstrates the \textit{"Sneaking"} pattern: an installation screen with fine print that installs third-party software unless explicitly disabled.

\begin{figure}[h]
\centering
\includegraphics[width=0.4\textwidth]{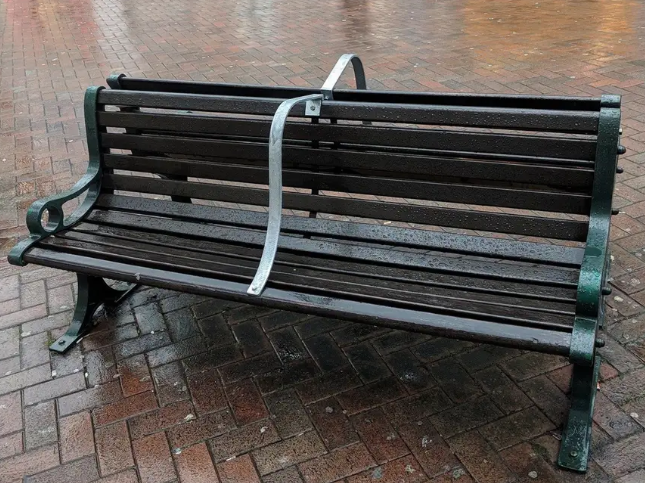}
\caption{Park bench with installed metal bar, example of physical deceptive pattern.}
\label{fig:banco}
\end{figure}

The unethical practices behind deceptive patterns have existed for decades, gaining attention from researchers in Computer Science, particularly in Human-Computer Interaction (HCI) \citep{MATHUR2021a}. Although explored primarily in digital technologies, deceptive patterns predate the internet and exist in the physical world. For example, companies offering free trial periods require credit card information and automatically charge customers without notice after the trial ends. Figure \ref{fig:banco} shows a bench with a metal bar preventing people from lying down or sitting together, favoring third-party interests over citizens' needs. This example is also considered as hostile design \citep{GONZATTO2017,SAVIC2014,ROSENBERGER2020}, linked to \textit{aporophobia} (rejection of the poor).

Another physical example includes mall escalators placed to force customers past storefronts. Thus, deceptive patterns represent a sociotechnical phenomenon: intentional design decisions in technical artifacts, digital or physical, benefiting third parties at users' expense. Their manifestation depends on technical aspects that operationalize them and social aspects that contextualize their use and meaning.

\subsection{The Semiotic Framework}
\label{sec:theframework}
Deceptive patterns challenge traditional design analysis by operating at the nexus of technology and human behavior, requiring a perspective that transcends isolated technical or social lenses. While approaches like behavioral psychology or ethical critiques highlight specific impacts -- such as cognitive manipulation or privacy violations -- existing literature often do not capture the interplay between intentional design choices and their societal consequences (Section \ref{sec:mappingresults}). 

According to danah boyd \citep{BOYD2008}, "The study of sociotechnical phenomena is about understanding the intersections between technologies and social practices." Deceptive patterns are implemented in technical artifacts and impact social life through unauthorized personal data collection \citep{BOSCH2016}, privacy invasion (\citep{LACEY2019,GREENBERG2014}), and emotional frustration \citep{CONTI2010,BOSCH2016}. Addressing these interconnected social and technical issues requires analysis from a sociotechnical perspective.

\begin{figure}[h]
    \centering
    \includegraphics[width=0.48\textwidth]{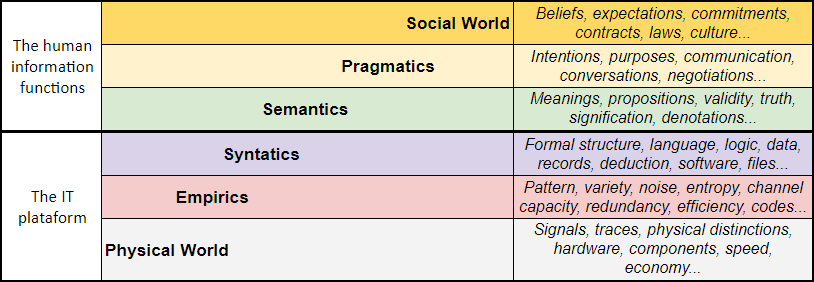}
    \caption{The six layers of the Semiotic Framework, by Ronald Stamper.}
    \label{fig:escadaS}
\end{figure}

The Semiotic Framework, developed by Ronald Stamper \citep{STAMPER1993} and rooted in Organizational Semiotics, is a promising tool for sociotechnical analysis, integrating Technical Platform and Human Functions -- see Figure \ref{fig:escadaS}. The Semiotic Framework has been successfully applied as a sociotechnical tool in interactive computing systems research \citep{hornung2015challenges,puska2020whatsapp,ferrari2020socially,BARONI2021,baranauskas2024socially}, making it well-suited for this work. Bellow are the six layers of the framework explained:

\begin{itemize}
    \item \textbf{Social World}: Addresses informational process impacts on social reality, beliefs, and actions; encompasses values, expectations, commitments, culture, and personal experiences.

    \item\textbf{Pragmatics}: Examines sign intentions, interpreter relationships, and contextual meaning within social settings.

    \item\textbf{Semantics}: Covers individual subjective interpretation of signs and meaning structures.

    \item\textbf{Syntactics}: Represents formal structures and patterns built with signs, including textual/visual construction and formal language.

    \item\textbf{Empirics}: Addresses sign coding, statistical properties, signal frequency, and transmission channel capacity.

    \item\textbf{Physical World}: Covers physical signal representation, including hardware, written materials, or smartphone screens, and physical properties like material and energy requirements.
\end{itemize}

\subsection{Related Works}

The authors \cite{nie2024shadows} presents a systematic mapping that resulted in the largest taxonomy of 64 deceptive pattern types, validated through industry feedback. The mapping reveals gaps in current detection tools, which identify only 50\% of patterns, and shows that existing datasets cover only half of these types. The findings emphasize the need for more comprehensive taxonomies, improved detection methods, and richer datasets to better address deceptive UI design.

\cite{gray2023mapping} presents a systematic mapping of research on deceptive patterns from 2014 to 2022, analyzing 79 empirical studies to identify prevalent contexts, methodologies, and disciplinary perspectives. The findings reveal diverse research approaches, with content analysis, experiments, and literature reviews being most common, while highlighting gaps such as limited cross-disciplinary integration and challenges in measuring the impact of deceptive patterns. The review offers insights and identifies future opportunities for methodological and conceptual advancements in understanding and regulating deceptive patterns.

\cite{chang2024theorizing} present a scoping review of 51 papers on deceptive patterns (2014-2023) that reveals significant undertheorization, with just over half (53.2\%) referencing any theory. Among 40 identified theories cited 99 times, Nudge theory dominated (28 references), followed by Dual Process theory (11 papers) and Choice Architect theory (6 papers), indicating heavy reliance on behavioral economics and psychology frameworks focused on user psychology while largely ignoring designer perspectives. Additionally, 46.5\% of theory references lacked sufficient detail or proper citations, and only 53\% were actually applied rather than merely mentioned. The authors call for enhanced theoretical incorporation and propose broader foundations spanning philosophy, psychology, sociology, and criminology to establish a more comprehensive base for future deceptive patterns research.

The authors \cite{santos2024they} examined the impact of Deceptive Patterns in digital games-design tactics that manipulate player behavior for profit-by developing a critical, educational game grounded in Self-Determination Theory (SDT). The primary contribution lies in designing and evaluating a mobile, interactive game that raises awareness among players and developers about the ethical implications and negative consequences of such patterns. The game employs a decision-making, reflective approach to foster critical understanding of how deceptive design choices undermine players’ psychological needs, such as autonomy, competence, and relatedness. The study's findings demonstrate that the game effectively promotes reflection and engagement, highlighting its potential as a tool for ethics-aware game design and player empowerment in the industry. The game proposed by the authors has an objective similar to our online catalog, although focused on deceptive patterns in games. The catalog presented in our work has a broad spectrum of patterns ranging from games to websites and other mobile apps. 

Authors \cite{castro2025teaching} present a study that engages with recent research on design patterns in digital platforms, especially within educational environments for children. It highlights the critical analysis of interfaces regarding the adoption of ethical (bright patterns) and manipulative (deceptive patterns) designs in the educational app Duolingo. This approach emphasizes balancing user engagement and autonomy, with a focus on children's digital rights and fostering reflection on responsible design practices. Thus, it supports the need for guidelines that promote transparency and protect against manipulative tactics in educational applications. 

The article by \cite{santos2026don} addresses the negative impact of Deceptive Patterns in digital games on player well-being, grounded in Self-Determination Theory to advocate the development of Radiant Patterns, which promote the satisfaction of basic psychological needs. It proposes an operationalized framework for these radiant patterns and presents the critical game The Good Dev, which uses decisions based on Radiant and Deceptive Patterns to foster players’ ethical reflection on healthy design, contributing to more ethical and player-centered digital environments. With a similar intention, the authors 

Unlike previous reviews that primarily focused on taxonomic expansion, methodological characterization, or theoretical enumeration, our study advances a sociotechnical understanding of deceptive patterns by integrating empirical evidence from a systematic mapping with an analytical interpretation grounded in the Semiotic Framework. Beyond cataloging the use or classifying pattern types, we examine how technical mechanisms and social norms interact to sustain deceptive design practices. This approach helps to bridge the gap between human-centered and technical analyses identified in prior works and offers a conceptual characterization of deceptive patterns as sociotechnical phenomena. The online catalog designed to support both user awareness and ethical design practice helps to exemplify specific patterns and their working mechanisms.

Building upon these theoretical and empirical foundations, the next section presents the methodological design adopted to investigate deceptive patterns as a sociotechnical phenomenon.



\section{Research Methodology}
\label{sec:resmethod}

\begin{figure}[h]
\centering
\includegraphics[width=0.48\textwidth]{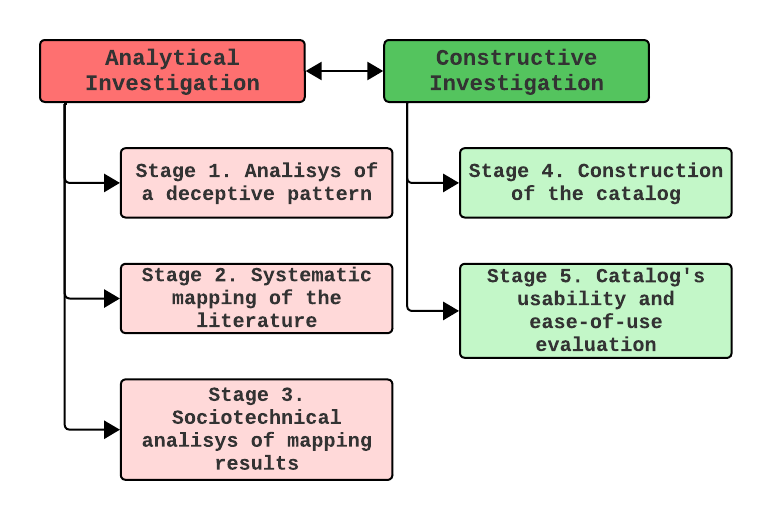}
\caption{Research methodology combining analytical and constructive investigations.}
\label{fig:metodologia }
\end{figure}

The methodology for this research is grounded in a conceptual-constructive approach, inspired by the framework proposed by \cite{oulasvirta2016hci} for understanding HCI research as a problem-solving process. This approach combines \textbf{analytical and constructive investigations} to achieve two main objectives: (1) understanding deceptive patterns through a sociotechnical lens and (2) developing a technical artifact to organize and present these patterns in an accessible manner. Figure \ref{fig:metodologia } provides a visual summary of the methodology used in the research.

The \textbf{analytical investigation} was conducted in three stages, each focusing on a specific aspect of deceptive patterns. These stages were designed to systematically dissect the deceptive patterns' nature, characteristics, and implications while considering their sociotechnical dimensions.

\emph{Stage 1: Analysis of a Deceptive Pattern}
The first stage focused on analyzing a single deceptive pattern known as the "Roach Motel" (a situation that is easy to enter but difficult to exit). The Semiotic Framework was used to examine the pattern across the six interconnected layers. This first analysis was important to confirm the usefullness of the framework on the study of deceptive patterns. The first author led the study, which was refined through iterative discussions with three experts on the Semiotic Framework, Deceptive Patterns, or both.  

\emph{Stage 2: Systematic Literature Mapping}
The second stage involved a systematic literature review conducted by the guidelines established by \cite{KITCHENHAM2007}. The first author developed a predefined protocol and validated it with two researchers with expertise in systematic reviews. The findings of the mapping were discussed with the same group of researchers involved in Stage 1, ensuring consistency and rigor. Details are further explained below.

\emph{Stage 3: Analysis of Sociotechnical Aspects}
The third stage involved scrutinizing the results of the systematic mapping to identify sociotechnical aspects-ranging from the physical to the social world-that were either acknowledged or overlooked in the literature. The Semiotic Framework served as the foundation for this analysis, which was conceptually validated by two researchers who participated in Stages 1 and 2.

The \textbf{constructive investigation} happened in two steps: the first step focused on developing a technical artifact: an interactive, self-explainable catalog of deceptive patterns identified in the scientific literature. The second step focused on evaluating the catalog's Usability and Easy of Use.

\emph{Stages 4 and 5: Construction and Evaluation of the Catalog}
During the development of the catalog, and in alignment with its objective of explaining what deceptive patterns are and how they operate, the idea emerged to incorporate practical examples. To this end, a subset of deceptive patterns was selected based on their feasibility of implementation. These examples were designed to provide users with a first-hand experiential understanding of how deceptive patterns function, while ensuring that participants were not exposed to harmful or manipulative real-world interfaces.
In addition to these interactive examples, the catalog includes all deceptive patterns identified in the systematic mapping study. The patterns are organized across eleven dedicated pages, each corresponding to a specific category of deceptive pattern.
Following the development phase, the catalog underwent an evaluation focusing on usability and ease of use, with a free exploration of the website by participants, the completion of a usability questionnaire, and a final focus group session, in which participants discussed their perceptions, experiences, and reflections about the catalog both among themselves and with the researchers.

The rest of the work is structured following the four especific objectives presented at the Introduction(Section \ref{sec:intro}): Section \ref{sec:goal1} presents the analysis and characterization of a deceptive pattern from a sociotechnical perspective and its results. Section \ref{sec:goal2} brings the systematic mapping of the literature on deceptive patterns and the results obtained. Section \ref{sec:goal3} presents the sociotechnical aspects considered and neglected in the literature and the results found. And last, but not least, Section \ref{sec:goal4} presents the online catalog, the construction process, and the evaluation results. 

\section{Goal 1: Characterizing Deceptive Patterns from a Sociotechnical Perspective }
\label{sec:goal1}




Addressing the first specific objective-analyze and characterize a deceptive pattern from a sociotechnical perspective-this section brings: first the analysis of the Roach Motel deceptive pattern, that served as a base for the characterization of deceptive patterns as a sociotechnical phenomena, presented next. 

\subsubsection{Analysis of the Roach Motel deceptive pattern}
\label{sec:roach}
\begin{figure}[h]
    \centering
    \includegraphics[width=0.48\textwidth]{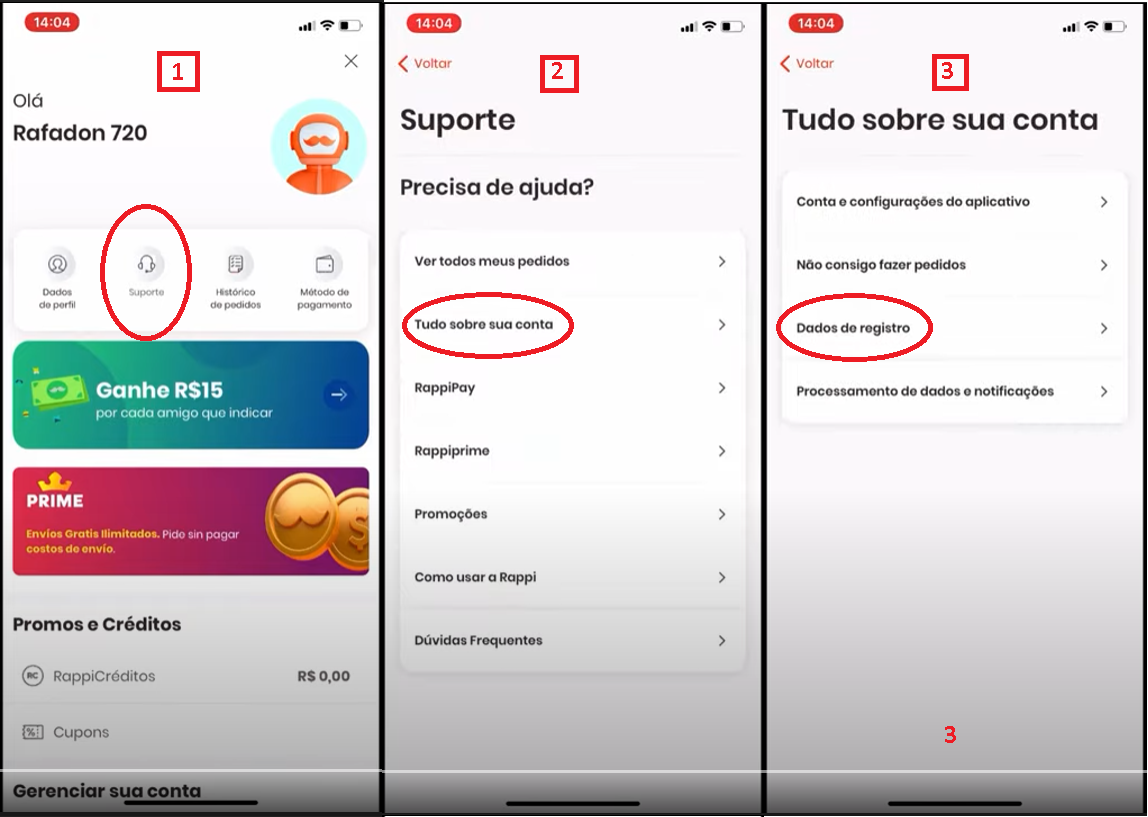}
    \caption{Example of the Roach Motel deceptive pattern.}
    \label{fig:hotel1}
\end{figure}

This section presents the analysis of a specific deceptive pattern known as the Roach Motel. The Roach Motel describes a situation that is easy to enter but difficult to exit, and it is commonly encountered when users attempt to delete an account or cancel a service. This deceptive pattern was selected due to its frequent occurrence across various online services and as a clear example of how a technical artifact can directly and negatively impact the social world.

Figures \ref{fig:hotel1} and \ref{fig:hotel2} illustrate an example of the Roach Motel during an account deletion process in an application, deliberately made complicated. By analyzing the example in Figures \ref{fig:hotel1} and \ref{fig:hotel2}, it is possible to observe that this deceptive pattern affects users in different but interconnected ways across the six layers of the Semiotic Framework.

To delete their account, the users must navigate through multiple screens: first clicking on "Support" (Screen 1), then "All About Your Account" (Screen 2), followed by "Registration Data" (Screen 3), and finally locating the option "Cancel My Account" (Screen 4). The term “Cancel” has an impact on the \textbf{semantic layer}: the users intends to close their account, possibly deleting their data, yet the application only offers the option to “Cancel” the account. The term "Cancel" does not carry the same meaning as "Delete" or "Remove." A common practice in such cases is for the company to retain users data even after an account closure request, creating a conflict of interest that affects the \textbf{pragmatic layer}.

\begin{figure}[h] 
    \centering 
    \includegraphics[width=0.48\textwidth]{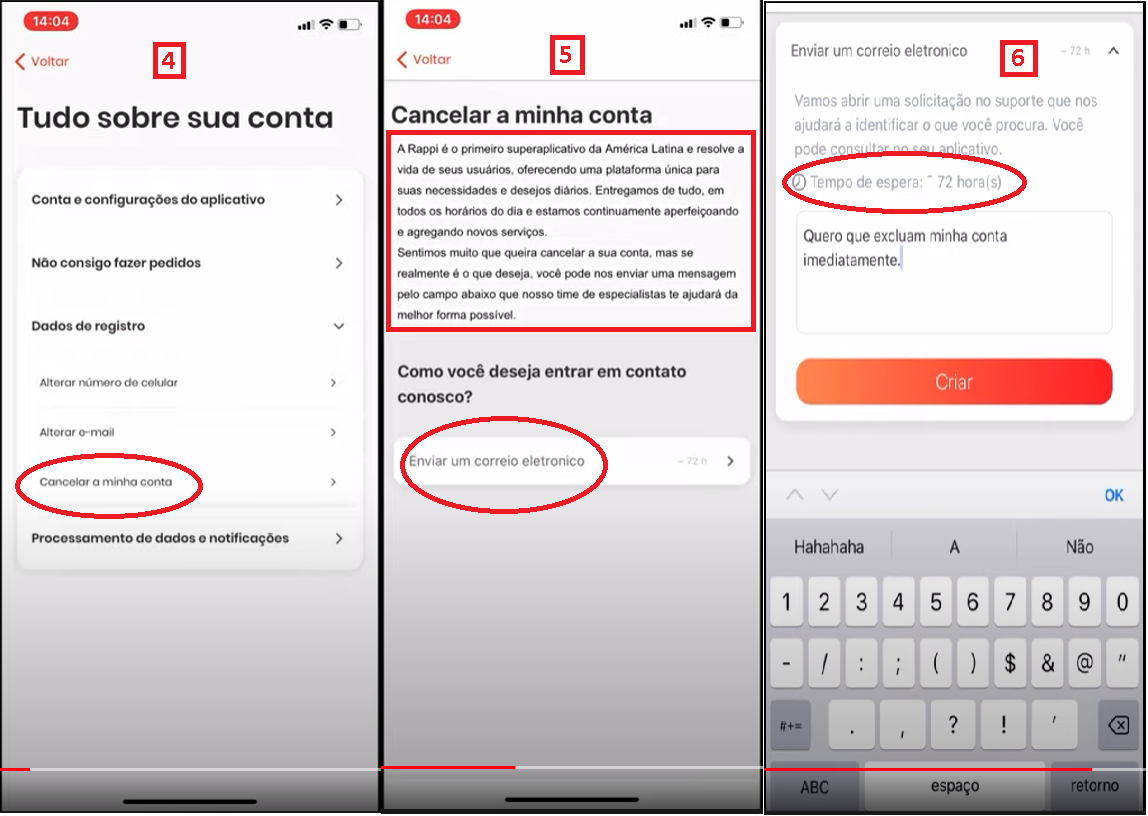} 
    \caption{Continuation of the Roach Motel deceptive pattern example.} \label{fig:hotel2} 
\end{figure}

Continuing the task, the users who believe they have finally found the option to delete their account encounters a lengthy message (Screen 5). The first half of this message contains irrelevant information, creating cognitive overload and affecting the \textbf{empirical layer}. The second half of the message employs emotionally charged language, which involves the \textbf{social world layer}: "We are very sorry that you want to cancel, but if this is truly your decision...". Additionally, the message informs the users that they must send an “email” to the application's support team to delete the account.

At this stage, the users are unable to complete the deletion independently, as they now rely on third-party intervention, impacting the \textbf{syntactic layer}. Finally, upon selecting the message submission option, the users gains the ability to take action (Screen 6); in this case, they must send an email to support, which may take up to 72 hours to process the request. This requirement affects the \textbf{physical world layer}, forcing the users to engage in an additional task (sending an email) and to wait an uncertain amount of time, potentially disrupting their plans.

The excessive number of screens, the volume of text, and the additional steps imposed on the users (having to send a message and wait for a support team) are all deliberately designed obstacles, impacting the \textbf{empirical layer}.

This analysis demonstrates that deceptive patterns influence user behavior by exploiting their intentions, knowledge level, and expectations and manipulating interface structures, text, imagery, and real-world constraints. Design decisions embedded in technical artifacts deny users the ability to make informed choices, leading them into undesirable situations. Furthermore, the example shows how the Semiotic Framework supports the analysis of deceptive patterns by considering both human and technical dimensions, allowing deceptive patterns to be characterized as a sociotechnical phenomenon.

The analysis was reviewed and discussed with three other researchers with prior knowledge of semiotics, two of whom had extensive experience with the framework, ensuring methodological rigor. The complete results of the analysis with the Semiotic framework were published in the XX Brazilian Symposium on Human Factors in Computing Systems (IHC 2021) \citep{BARONI2021}. The study reveals nuances often overlooked in purely technical assessments, such as the social and emotional consequences of misleading or inconveniencing users.

\subsubsection{Characterizing Deceptive Patterns with the Semiotic Framework}
\label{sec:characterization}

Deceptive patterns aim to influence user behavior in favor of other stakeholders by exploiting both technical and human aspects of interaction. Computational systems shape social behaviors by promoting certain actions while inhibiting others \citep{FOGG1999}. A deceptive pattern becomes truly \textit{deceptive} when it exploits elements at any framework layer against users. However, these patterns often leverage multiple layers in an interconnected manner. Analyzing deceptive patterns through the Semiotic Framework allows for a structured understanding of their mechanisms and their impact on users.

In the \textbf{Social World}, deceptive patterns manipulate user behaviors, emotions, and values. Designers may exploit personal characteristics such as age, beliefs, and social context to craft deceptive interfaces. This layer highlights the tangible consequences of deceptive patterns, from emotional distress \citep{MAIER2020,GRAY2020} to financial loss. Techniques like ConfirmShaming \citep{BRIGNULL2023} and Roach Motel \citep{MATHUR2019} take advantage of users' social and cognitive vulnerabilities.

At the \textbf{Pragmatic Layer}, deceptive patterns distort users' intentions and expectations. Users anticipate predictable system responses, but deceptive designs redirect their actions for the benefit of other stakeholders. Examples include disguised download buttons \citep{BRIGNULL2023} and artificial urgency tactics like Scarcity \citep{MATHUR2019}.

The \textbf{Semantic Layer} involves the manipulation of meaning in interface elements-buttons, text, images, and colors-to deceive users. Misleading language and ambiguous consent mechanisms exploit users’ limited technical knowledge, as seen in vague cookie consent notices \citep{UTZ2019} and misleading questions \citep{MATHUR2019,BRIGNULL2023,de2024dark,de2023dark}.

At the \textbf{Syntactic Layer}, deceptive patterns exploit interface structure and task flows. Elements such as button placement \citep{lima2022estrategias}, interface interference \citep{GRAY2018}, and forced actions manipulate user interactions \citep{albuquerque2024investigating}. An example is swapping the positions of confirm and cancel buttons or forcing users to watch ads before continuing their tasks.

The \textbf{Empirical Layer} addresses statistical properties of deceptive pattern elements, such as excessive text, disruptive animations, or delays in essential features. Information overload, misdirection \citep{BRIGNULL2023}, and obstruction tactics \citep{GRAY2018} increase cognitive load and hinder users' ability to make informed decisions.

Finally, the \textbf{Physical World} layer includes constraints related to hardware and physical interactions. Deceptive patterns may impose additional user effort, such as requiring phone calls or physical letters to cancel services. Interface manipulations on touchscreen devices or restrictive hardware policies compel users to make unintended choices.

By analyzing deceptive patterns through these six interconnected layers, a sociotechnical perspective emerges. This layered approach provides a comprehensive understanding of deceptive patterns, uncovering critical elements that would otherwise be overlooked in isolated analyses. Addressing deceptive patterns requires a multifaceted strategy, integrating social, technical, and physical dimensions to mitigate their effects effectively. For example: a cookie consent banner that uses the term "\textit{Enhance your experience}" instead of explicitly stating "\textit{We collect and share your data}" could mislead users. If only the syntactic layout is considered, the deceptive language (semantic layer) might go unnoticed.

\section{Goal 2: Mapping the Literature on Deceptive Patterns}
\label{sec:goal2}

Addressing the second specific objective-to systematically map the literature about deceptive patterns-below are presented the protocol used for the systematic mapping, followed by the results obtained with the mapping. 

\subsection{Systematic Mapping Protocol - Method}
\label{ssec:protocol}


The systematic literature mapping was planned and conducted based on the guidelines proposed by \cite{KITCHENHAM2007}. A search string was designed and applied to digital libraries to retrieve a set of scientific publications relevant to the mapping objectives. These publications underwent a multi-stage selection process to filter out unrelated works. Finally, data were extracted from the selected studies for analysis, ensuring scientific rigor and reproducibility.

\subsubsection{Objective and Research Questions}

\begin{table}[]
\centering
\begin{tabular}{|l|l|}
\hline
\textbf{Analyze}             & Scientific publications                \\ \hline
\textbf{With the purpose of}   & \begin{tabular}[c]{@{}l@{}}Identifying and cataloging \\ deceptive patterns  \end{tabular}\\ \hline
\textbf{In relation to} & \begin{tabular}[c]{@{}l@{}}Types of deceptive patterns \\ and possible categories, \\
characteristics, and methods \\ for identification\end{tabular} \\ \hline
\textbf{From the perspective of} & \begin{tabular}[c]{@{}l@{}}Computer Science\\ researchers\end{tabular} \\ \hline
\textbf{In the context of} & User interfaces                  \\ \hline
\end{tabular}
\caption{Objective of the systematic mapping study.}
\label{tab:mapping_objective}
\end{table}

The \textbf{primary objective} of this mapping study, formulated using the Goal-Question-Metric (GQM) paradigm \citep{BASILI1988}, (Table \ref{tab:mapping_objective}) was to identify and analyze scientific research on deceptive patterns within the field of Computer Science. The specific objectives were:

\begin{itemize}
    \item 1. Identify and catalog deceptive patterns studied or presented in the Computer Science literature.
    \item 2. Analyze how researchers have investigated and classified deceptive patterns.
\end{itemize}

The main \textbf{research question}: "\textit{What is the state of the art on deceptive patterns in the Computer Science literature?}" The definition of deceptive patterns adopted in this study follows \cite{BOSCH2016}: "... an established solution for exploiting and deceiving users in a generic form". This broad definition ensured that the study would cover as many deceptive patterns as possible. Table \ref{tab:research_questions} shows the sub-questions formulated to help answer the main research question.

\begin{table}[h]
    \renewcommand{\arraystretch}{1.2} 
    \begin{tabular}{|p{0.45\textwidth}|} 
        \hline
        \textbf{Research Questions} \\
        \hline
        \textbf{RQ1 - Which deceptive patterns have been studied or presented?} \\ 
        To identify deceptive patterns reported in scientific literature and gain a deeper understanding of the state of the art in deceptive patterns within Computer Science. \\
        \hline
        \textbf{RQ2 - What application domains do the studies focus on in relation to deceptive patterns?} \\ 
        To understand if there is a correlation between domains and the presence of deceptive patterns (e.g., websites, online shopping, advertisements) and compare their usage across different domains. \\
        \hline
        \textbf{RQ3 - How are deceptive patterns defined?} \\ 
        To analyze how researchers define deceptive patterns and compare these definitions to derive a general definition that can aid in their identification. \\
        \hline
        \textbf{RQ4 - How are deceptive patterns being cataloged?} \\ 
        To determine whether there are existing classification schemes for deceptive patterns, how they work, and whether they can be generalized. \\
        \hline
        \textbf{RQ5 - What theoretical foundations or methodological approaches are used in the studies?} \\ 
        To identify theoretical and methodological frameworks that support deceptive pattern research. \\
        \hline
        \textbf{RQ6 - What research problems are being addressed?} \\ 
        To understand the context in which deceptive patterns are studied and identify research gaps and unresolved issues. \\
        \hline
        \textbf{RQ7 - What are the key contributions of the studies?} \\ 
        To assess which aspects researchers are focusing on and identify research gaps. \\
        \hline
        \textbf{RQ8 - Are there any forms of evaluation or experimentation?} \\ 
        To identify the evaluation approaches used by researchers in deceptive pattern studies. \\
        \hline
    \end{tabular}
    \caption{Research Questions}
    \label{tab:research_questions}
\end{table}

\subsubsection{Scope and Language}

Constructing a well-defined search string ensures representative results from digital library search engines. The search string for the systematic mapping was built based on \citep{KITCHENHAM2007}. So as to ensure total coverage of the literature, the initial search string used on Google Search contained only the phrase “dark patterns” (enclosed in quotation marks). The term "dark patterns" was still widely used at the time. Refer to \citep{} for further details.


The default sorting filter in Google Scholar was maintained, meaning the results were listed according to their relevance at that moment. The first ten publications that appeared in the search were selected \citep{ZAGAL2013,BOSCH2016,GRAY2018,MATHUR2019,NOUWENS2020,GREENBERG2014,TRICE2018,CHROMIK2019,LACEY2019,CONTI2010}. The keywords from these ten publications were analyzed to identify relevant terms, which were then used to construct the following search string:

\bigskip
\noindent\fbox{%
    \parbox{8.25cm}{%
        "dark pattern" OR "dark patterns" OR "malicious interface" OR "evil interface" OR "dark design pattern" OR "dark design patterns" OR "dark side" OR "dark strategies" OR "dark strategy" OR “dark interface”
    }%
}
\bigskip

Subsequently, the term "deceptive pattern" was added to the search string, and the results were compared to those obtained without its inclusion. The analysis confirmed that no relevant articles had been omitted due to the absence of this term at the time of the search.

The following digital libraries were selected for the systematic mapping, as they are among the most prominent and well-recognized repositories in the field of Computer Science:

\begin{itemize}
    \item ACM Digital Library;
    \item IEEExplore;
    \item Scopus (search limited to Computer Science).
\end{itemize}

English was chosen as the language for the systematic mapping, as it is the primary language used in international publications in Computer Science.

\subsubsection{Strategy for Article Selection}

The scientific publications retrieved from the digital libraries were selected through a two-stage filtering process, where inclusion and exclusion criteria were applied. The first filter involved analyzing the title and abstract of each publication. The second filter consisted of reading the introduction and conclusion sections of the publications that passed the first filter. If uncertainties remained after the second filter, the entire publication was read to ensure its inclusion or exclusion. Finally, all publications that met the inclusion criteria after both filters proceeded to the data extraction phase.

The first and second filters were conducted by two researchers with prior experience in systematic mappings to reduce selection bias. The Cohen-Kappa coefficient \citep{LANDIS1977} was used to measure agreement between the researchers, resulting in a coefficient of 0.8183 (indicating an Almost Perfect agreement). The inclusion and exclusion criteria applied during the selection process are detailed below:

\textbf{Inclusion Criteria:}
\begin{itemize}
    \item \textbf{IC1:} Scientific publications related to deceptive patterns in Computer Science.
    \item \textbf{IC2:} Articles that explicitly address deceptive patterns in Computer Science.
\end{itemize}

\textbf{Exclusion Criteria:}
\begin{itemize}
    \item \textbf{EC1:} The content does not directly address deceptive patterns in Computer Science.
    \item \textbf{EC2:} The article is not written in English.
    \item \textbf{EC3:} The article is duplicated, appearing in multiple digital libraries.
    \item \textbf{EC4:} If an article has an extended version, only the most comprehensive version is selected.
    \item \textbf{EC5:} The full text of the article is not accessible or requires payment.
\end{itemize}

\subsubsection{Data Extraction and Coding Procedures}
\label{ssec:data}

All articles that passed the first and second filtering stages were read in full to ensure they aligned with the scope of the study. Data extraction was then performed based on previously defined research sub-questions. Each article received a unique ID, and metadata were extracted from all the articles (listed below). Once extracted, all the data collected was reviewed and analyzed.

\begin{itemize}
\item \textbf{MD1}: Title of the publication;
\item \textbf{MD2}: Year of publication;
\item \textbf{MD3}: Authors;
\item \textbf{MD4}: Authors' affiliated institution;
\item \textbf{MD5}: Type of publication (Journal, Event);
\item \textbf{MD6}: Publication venue.
\end{itemize}


The extraction process followed the Data Extraction Table (itemized below), which outlines the specific values to be retrieved from the article text. Each research sub-question is associated with one or more data points. Some data points required inference by the researchers if not explicitly stated by the authors, such as Data ID 10 (Application Domain) and Data ID 16 (Type of Contribution). Table \ref{tab:extraction} shows the extracted data points.

\begin{table}[h]
    \centering
    \resizebox{0.48\textwidth}{!}{ 
        \begin{tabular}{|c|l|c|}
            \hline
            \textbf{ID} & \textbf{Data} & \textbf{QP} \\ \hline
            1  & Unique publication ID & - \\ \hline
            2  & Publication title & MD1 \\ \hline
            3  & Year of publication & MD2 \\ \hline
            4  & Authors & MD3 \\ \hline
            5  & Authors' institution & MD4 \\ \hline
            6  & Type of publication & MD5 \\ \hline
            7  & Publication venue & MD6 \\ \hline
            8  & Deceptive pattern name & RQ1 \\ \hline
            9  & Deceptive pattern description & RQ1 \\ \hline
            10 & Application domain & RQ2 \\ \hline
            11 & Definition of deceptive pattern & RQ3 \\ \hline
            12 & Cataloging method & RQ4 \\ \hline
            13 & Theoretical foundations & RQ5 \\ \hline
            14 & Addressed problem & RQ6 \\ \hline
            15 & Main contribution & RQ7 \\ \hline
            16 & Type of contribution & RQ7 \\ \hline
            17 & Evaluation method & RQ8 \\ \hline
            18 & Evaluation metrics & RQ8 \\ \hline
        \end{tabular}
    }
    \caption{Data extraction table.}
    \label{tab:extraction}
\end{table}

This structured extraction strategy ensures the collected data is comprehensive and aligned with the research objectives.

\subsubsection{Strategy for Data Analysis}
After the extraction phase, the data from the articles were organized to address each research question in the mapping. To answer \textbf{RQ1} (Which deceptive patterns have been studied/presented?), all deceptive patterns identified in the selected articles were listed along with their respective explanations. The deceptive patterns were then categorized and grouped based on their similarities, and where applicable, the categories from \cite{GRAY2018} and \cite{MATHUR2019} were used.

To answer \textbf{RQ2} (What are the application domains that the articles focus on in relation to deceptive patterns?), the application domains associated with the deceptive patterns mentioned in each article were identified. Subsequently, the results were categorized and grouped: similar and synonymous themes were replaced to make them equivalent. For example, "buying online" was grouped with "shopping websites," and the frequency of each domain's appearance was counted.

To answer \textbf{RQ3} (How are these deceptive patterns defined?), the extracted definitions were examined for the adjectives and verbs used by the authors to describe the deceptive patterns, such as "deceptive" or "trick." Additionally, the definitions were categorized as Unique, from \cite{GRAY2018} or \cite{BRIGNULL2023}.

To address \textbf{RQ4} (How are these deceptive patterns being cataloged?), \textbf{RQ5} (What are the theoretical foundations of the publications?), and \textbf{RQ6} (What is the research problem being studied?), the content extracted from the articles was summarized and categorized into six different types of problems. This categorization was created through open coding based on the themes of the problems discussed in the articles.

To answer \textbf{RQ7} (What is the main contribution of the article?), the process was similar to that of RQ2. For \textbf{RQ8} (Is there any form of evaluation or experiment?), the articles that presented some form of evaluation or experiment were categorized into three types: Experiments with users, designers, or students. The data analysis was conducted as described above, and another researcher reviewed the results.

The metadata were grouped and counted: the number of publications on deceptive patterns over the years, the type of these publications (Event or Journal), where the articles were published (name of the event or journal), institutions that published on deceptive patterns, and the countries of these institutions.

\subsection{Systematic Mapping Results}
\label{sec:mappingresults}

The systematic mapping was conducted in the main computing databases (ACM, IEEE, Scopus), providing a rigorous and up-to-date literature overview. Additionally, the sociotechnical aspects of the articles found in the mapping were analyzed. The search included all the articles published up until July of 2022 (date of 07/07/2022). The amount of articles returned in the search and accepted in each filter are presented in Table \ref{tab:numerosArtigos}. A list containing the names of all (97) articles selected in the mapping can be found in the following link: \href{https://docs.google.com/spreadsheets/d/178Ce3HrhzRiUocvI5JiAhsjFdM0hYrP04kn6pVNW8\_0/edit?usp=sharing}{\textcolor{blue}{\textbf{\underline{Selected Articles}}}}. 

\begin{table}[]
\centering
\begin{tabular}{|l|l|}
\hline 
\textbf{Stage}         & \textbf{Quantity} \\ \hline
Returned Initial Search     & 1655           \\ \hline
Accepted in First Filter    & 240           \\ \hline
Accepted in Second Filter   & 97           \\ \hline
Repeated                    & 181           \\ \hline
Rejecteds                   & 1377           \\ \hline
\end{tabular}
\caption{Number of articles returned in each step of the systematic mapping.}
\label{tab:numerosArtigos}
\end{table}

\begin{figure}[h]
    \centering
    \includegraphics[width=0.48\textwidth]{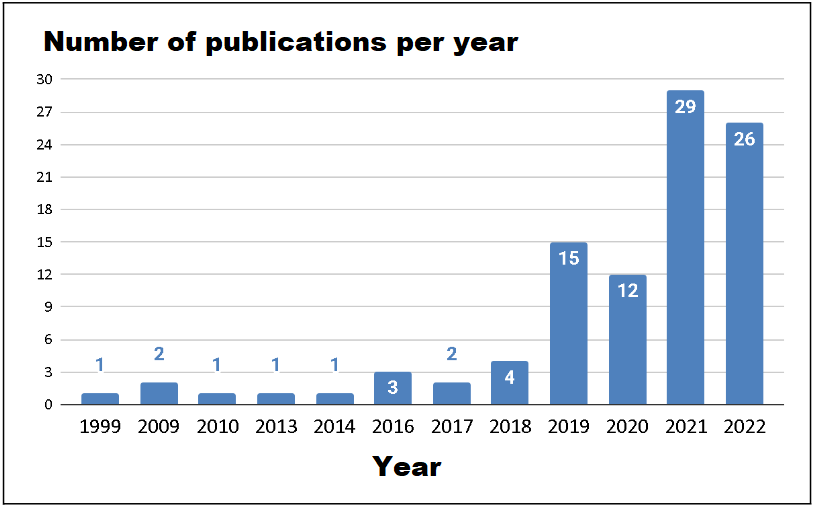}
    \caption{Distribution of articles over the years.}
    \label{fig:years}
\end{figure}

Figure \ref{fig:years} shows the distribution of articles by publication year. In 1999, Fogg's work on persuasive technologies was published \citep{FOGG1999}. In 2010, the term "dark patterns" was coined by Harry Brignull \citep{BRIGNULL2023}. In 2018, the term gained popularity in academia after the publication of \citep{GRAY2018}, leading to increased research on the topic. 

\begin{figure}[h]
    \centering
    \includegraphics[width=0.48\textwidth]{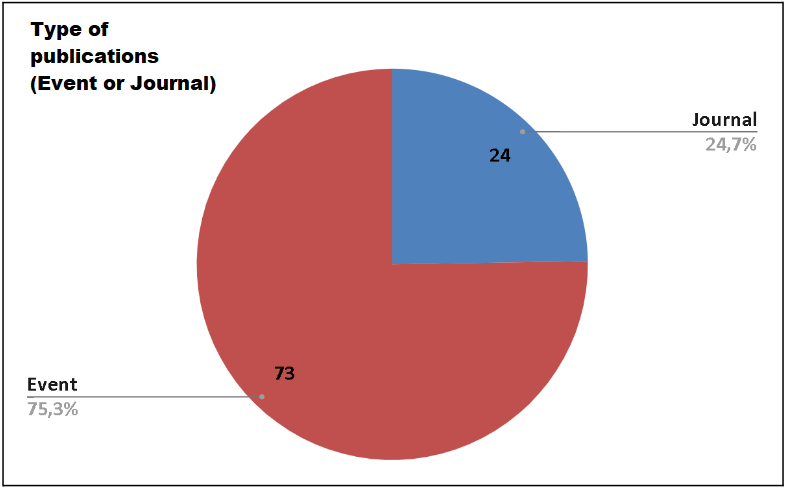}
    \caption{Types of publications.}
    \label{fig:types}
\end{figure}

\begin{figure}[h]
    \centering
    \includegraphics[width=0.48\textwidth]{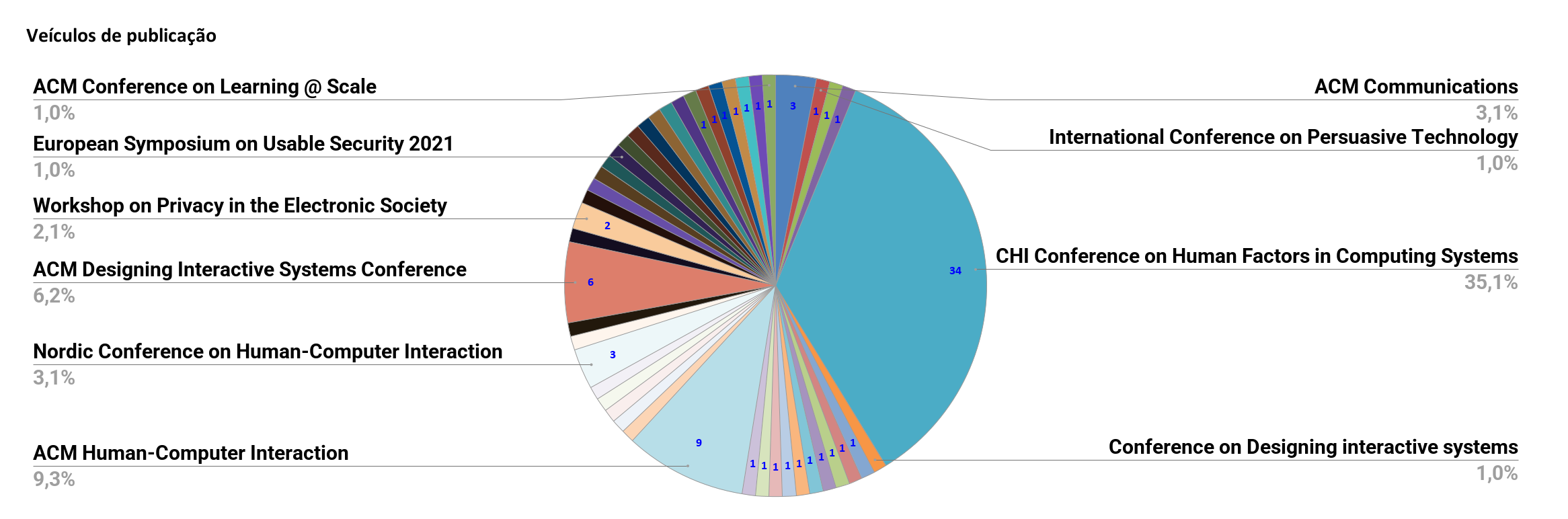}
    \caption{Publication venues.}
    \label{fig:venues}
\end{figure}

Figure \ref{fig:types} shows the quantities of the two main types of publications: Journals and Events. Twenty-four articles were published in journals, and the remaining 73 were published in events. Figure \ref{fig:venues} presents the different publication venues where articles were published. In total, 46 different publication venues were identified. The ACM CHI Conference on Human Factors in Computing Systems is the venue that publishes the most work on deceptive patterns and interfaces: 34 out of the 97 selected articles, followed by the ACM Human-Computer Interaction journal, which has nine published articles.


\begin{table}[]
\centering
\begin{tabular}{|l|l|}
\hline 
\textbf{Categories}         & \textbf{Quantity} \\ \hline
Disorientation              & 25           \\ \hline
Interface Interference      & 20           \\ \hline
Hiding                      & 15           \\ \hline
Social Pressure             & 15           \\ \hline
Obstruction                 & 14           \\ \hline
Forced Action               & 12           \\ \hline
Monetary                    & 8            \\ \hline
Without Consent             & 7            \\ \hline
Delay                       & 4            \\ \hline
Surveillance Capitalism     & 3            \\ \hline
Temporal                    & 3            \\ \hline
\end{tabular}
\caption{Categories of deceptive patterns and the number of types.}
\label{tab:categories}
\end{table}

\textbf{Research Question (RQ) 1: Which deceptive patterns have been studied/presented?} In total, 126 types of deceptive patterns were identified and divided into 11 categories (Table \ref{tab:categories}). These patterns can be seen at the following link: \href{https://docs.google.com/spreadsheets/d/178Ce3HrhzRiUocvI5JiAhsjFdM0hYrP04kn6pVNW8\_0/edit?usp=sharing}{\textcolor{blue}{\textbf{\underline{List of deceptive patterns found}}}}. The patterns types are unevenly distributed among the categories: the Disorientation category has the most types (25), while the Temporal and Surveillance Capitalism categories have only three types each. Of the 126 types identified, 26 appeared in more than one article simultaneously.


From the 97 articles analyzed, 34 presented some form of "established solution to exploit and deceive users in a generic manner" \citep{BOSCH2016} (the definition of deceptive patterns used throughout the mapping) or categories of such solutions intended to manipulate users and their behavior. In other words, just over a third of the mapped articles introduce new types, and among these, Gray et al. contributed with 12 unique types, followed by \cite{CONTI2009} with 11 unique types, and \cite{MATHUR2019} with 10. In total, eight groups of authors are responsible for more than half (70) of the unique types identified.

All deceptive patterns mentioned in the articles were compared, and duplicates or similar patterns were grouped into the same types. The list of all deceptive patterns can be found in the online catalog.

\begin{figure}[h]
    \centering
    \includegraphics[width=0.48\textwidth]{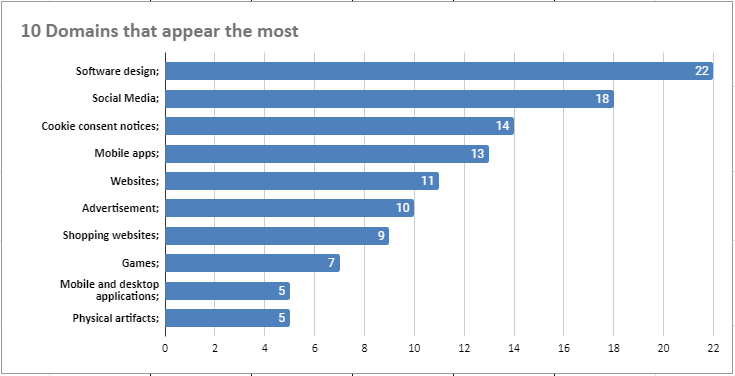}
    \caption{Most frequent application domains.}
    \label{fig:domains}
\end{figure}

\textbf{RQ2: To which application domains are these deceptive patterns related?} The most frequent application domain in the articles was "Software Design," appearing in 22 articles, followed by "Social Media" in 18 articles, and "Cookie Notices" in 14 articles. These works primarily focus on the software design process, its interfaces, and designers' decisions during the design process. In total, 30 domains were identified, reflecting a high diversity of application domains, reinforcing the idea that deceptive patterns can appear anywhere there is a user interface. Figure \ref{fig:domains} shows the top 10 most frequent domains. 

\begin{figure}[h]
    \centering
    \includegraphics[width=0.48\textwidth]{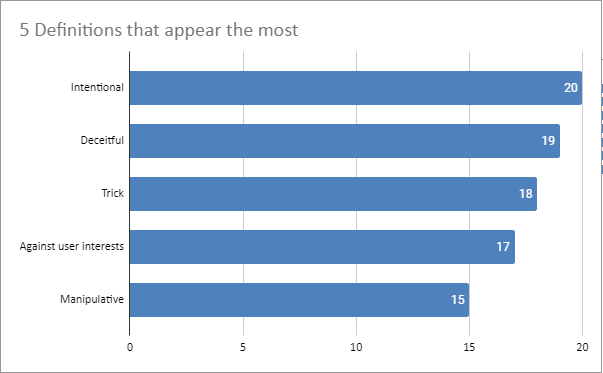}
    \caption{Count of definition types found.}
    \label{fig:definitions}
\end{figure}


\textbf{RQ3: How are these deceptive patterns defined?} Forty-seven distinct/unique definitions were found in the articles. Most articles explicitly mention the intentionality of deceptive patterns in their definitions (20). Additionally, many articles define deceptive patterns as "Deceptive" (19), "Tricks" (18), or "Acting against user interests" (17). There are also at least twenty other different but non-exclusive definitions. In 22 articles, the authors only mention or discuss deceptive patterns without providing or referencing any definition. Figure \ref{fig:definitions} shows a count of the different types of definitions found. Of the 97 mapped articles, almost half (47) provide unique definitions, and only 28 articles base their definitions on established authors in the field \citep{GRAY2018,MATHUR2019,BRIGNULL2023}. The results indicate a lack of a clear and consistent conceptual foundation among researchers, confirming the findings of \cite{MATHUR2021a}. Only 20 articles mention the intentionality of patterns in their definitions, which may indicate a lack of shared understanding about the fundamental aspects of deceptive patterns among researchers. Intentionality is a crucial element of deceptive patterns, and without a common conceptual framework that includes this aspect, there is potential for misunderstandings in the field. 

\begin{figure}[h]
    \centering
    \includegraphics[width=0.48\textwidth]{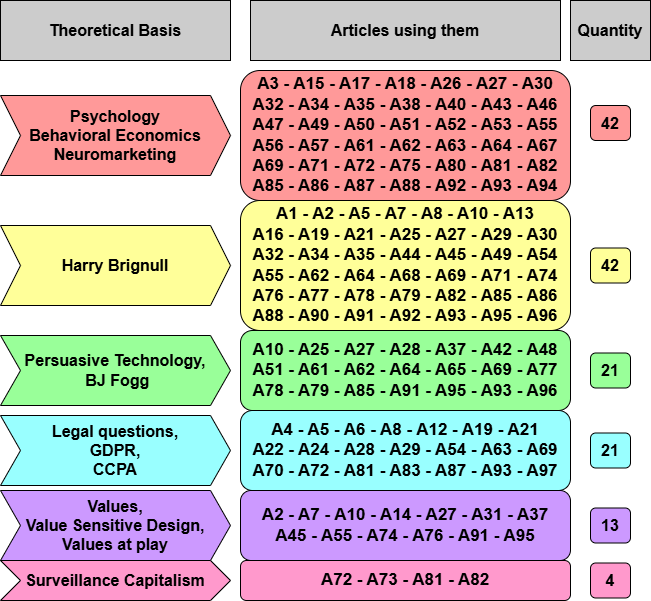}
    \caption{Types of references used in the articles.}
    \label{fig:references}
\end{figure}

\textbf{RQ4: How are these deceptive patterns being cataloged?} Only 23 articles presented some form of cataloging or explanation of collecting deceptive patterns. Overall, a few works (3) catalog types of deceptive patterns, while in most (18) works, deceptive patterns are collected and analyzed manually and then categorized by researchers using existing categorizations, such as those by \cite{BRIGNULL2023} or \cite{GRAY2018}. In one work (\citep{MATHUR2019}), the authors used web crawlers to collect data from the internet automatically but still analyzed it manually afterward. The three articles \citep{CONTI2010,NYSTROM2020,UTZ2019} that present cataloging methods: in one article \citep{CONTI2010}, researchers cataloged deceptive patterns with the help of participants in different experiments; in another \citep{NYSTROM2020}, researchers introduced a new framework called the Visibility-Darkness matrix, based on the Intention-Outcome matrix. The last article \citep{UTZ2019} presents a tool that automatically extracts information from websites, including cookie notice information, facilitating automated analysis. 

\textbf{RQ5: What are the theoretical foundations of the publications?} Figure \ref{fig:references} shows the different types of references used in the articles. Analyzing the theoretical foundations of the studies reveals a wide variety of sources on which the authors base their work. The original work by \cite{BRIGNULL2023} was cited by 42 articles, and B.J. Fogg's work on "persuasive technologies" \citep{FOGG1999} was cited by 21 articles. Almost half of the articles (42) draw on psychology and related fields, such as cognitive science, behavioral economics, and neuromarketing. For example, Kahneman's Dual Process Theory \citep{KAHNEMAN2003}, Bem's Self-Perception Theory \citep{BEM1967}, or the Fear of Missing Out (FoMO) \citep{PRZYBYLSKI2013}. Twenty-one works are based on legal articles or related to GDPR (General Data Protection Regulation in Europe) \citep{GDPR} or CCPA (California Consumer Privacy Act) \citep{CCPA}. Seventeen works are grounded explicitly in economics. Finally, 13 articles are based on discussions about values (of users and designers), such as Value Sensitive Design (VSD) \citep{FRIEDMAN2008,FRIEDMAN2019}. 

\textbf{RQ6: What are the research problems studied?} Seven main categories of problems addressed in the articles were identified, ranging from ethical and privacy issues to consequences and surveillance:
\begin{enumerate}
    \item \textit{Ethics}: Articles investigating the relationships and impacts of deceptive patterns on ethical issues, such as how ethics are used or inform designers' daily work, how ethics are discussed on social media, and the lack of research describing designers' ethical engagement.
    \item User Consent: Research addressing how deceptive patterns are applied to induce or manipulate user consent to options that benefit parties other than the users.
    \item Privacy: Problems describing the effects of deceptive patterns on user privacy. For example, privacy in the context of location tracking and assisting users in making privacy-related decisions.
    \item Behavioral Issues: Research focused on problems related to manipulating user behavior affected by deceptive patterns. For example, how persuasive technologies can be used for manipulation, how designers lack tools to measure how their choices shape behavior, and how researchers have not yet provided the necessary tools to combat the influences of deceptive patterns.
    \item Understanding and Mapping: Problems related to the concept, understanding, or cataloging of deceptive patterns. For example, factors influencing users to become victims of deceptive patterns, how many deceptive patterns appear in mobile applications, and how the cute design of domestic robots can create an ambivalent power relationship between the users and the robot.
    \item Consequences: Works concerned with the consequences of specific deceptive patterns and design decisions in general on users and society. For example, there is a lack of tools to understand these design (personal, social, moral, ethical, economic) consequences, and few studies have explored the adverse effects of persuasive technologies.
    \item Surveillance Capitalism: Works focusing on characteristics and problems directly related to surveillance capitalism \citep{ZUBOFF2019}, a business model that predicts and influences user behavior to maximize advertising value.
\end{enumerate}

\begin{figure}[h]
    \centering
    \includegraphics[width=0.48\textwidth]{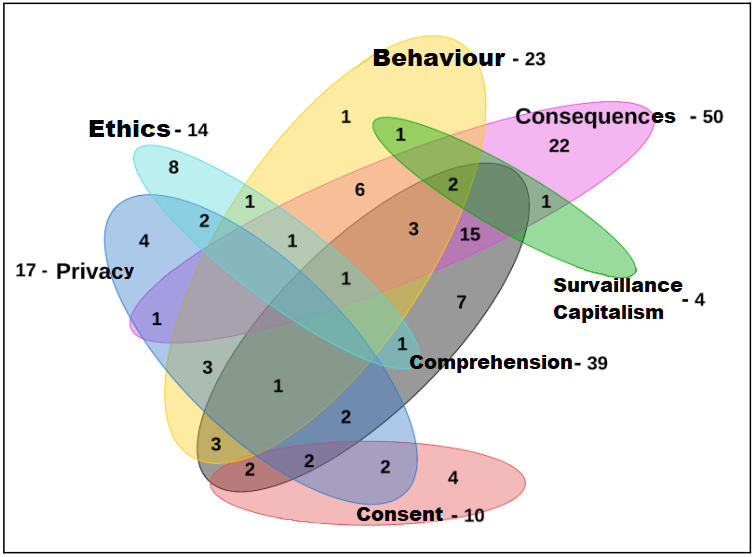}
    \caption{Venn diagram representing the problem categories.}
    \label{fig:problems}
\end{figure}

Most articles address problems in more than one category. Figure \ref{fig:problems} shows a Venn diagram with the number of articles in each category and their intersections: 50 articles address the Consequences of design, which are problems related to the outcomes and impacts of design choices and deceptive patterns on users or society in different usage scenarios and interfaces; 39 articles discuss problems associated with Understanding and Mapping deceptive patterns, which are problems related to the description and understanding of specific deceptive patterns and different categories; 23 articles address Behavioral Issues of users, mainly problems related to the manipulation of user behavior affected by deceptive patterns; 17 articles discuss Privacy problems, which describe the effects of deceptive patterns on user privacy; 14 articles address Ethical issues, investigating the relationships and impacts of deceptive patterns on ethical matters; 10 articles discuss Consent, addressing how deceptive patterns are applied to induce or manipulate user consent to options that benefit parties other than the users; and 4 articles deal with problems related to Surveillance Capitalism, a concept in political economy that denotes the widespread collection and commodification of personal data by corporations.

\textbf{RQ7: What is the main contribution of the articles?} Articles presented various types of contributions, some presenting more than one type: 36 articles present different types of experiments, primarily with users but also with students and professional designers. Twenty-nine articles also show various types of deceptive interfaces or categorizations; 12 articles include discussions focused on ethics, such as ethical issues in persuasive computing, the ethical responsibilities of professionals, and user discussions about ethical issues surrounding deceptive interfaces; 16 articles discuss cookie banners, consent notices, and legal issues \citep{GDPR,CCPA}. Examples include how the design of these notices impacts user behavior and consent and how and whether websites and applications comply with recent user consent laws \citep{GDPR,CCPA}. There are also examples of how deceptive patterns violate users' legal rights and how manipulation can be used for good and evil. Discussions on "surveillance capitalism" and an analysis of the origins of deceptive patterns are also included.

\begin{figure}[h]
    \centering
    \includegraphics[width=0.48\textwidth]{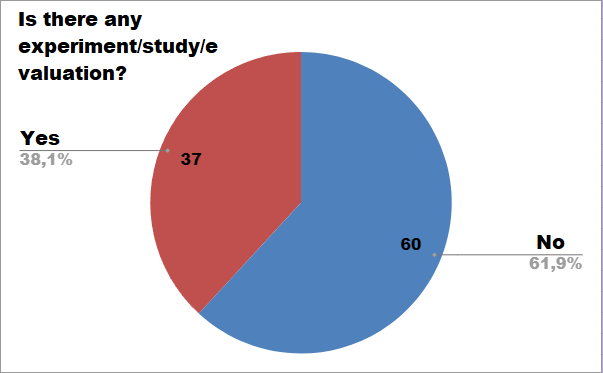}
    \caption{Number of works presenting some study/evaluation/experiment.}
    \label{fig:experiments}
\end{figure}

\textbf{RQ8: Is there any form of evaluation or experiment?} Thirty-seven (37) articles presented some form of evaluation or study (Figure \ref{fig:experiments}), and overall, the experiments have been well-defined and documented. 
The articles document observations and questionnaires on user behavior in various situations involving deceptive interfaces and deceptive patterns (such as browsing the web and interacting with cookie notices, mobile games, or social media). The objectives vary but share a common purpose: to understand the effects of deceptive design patterns (real or created by researchers) in different interfaces (cookie banners, websites, social media) on user behavior, feelings, and consent. Studies with students and professionals focused more on the role of the designer: the ethical values and intentions of professionals during the creation and development of interfaces for websites or consent notices. There are also two studies that focus on evaluations: an evaluation of guidelines, aimed at professionals, for designing applications that promote good user behavior\citep{CHAUDHRY2021}; and an evaluation of manuals (aimed at professionals) for the use and implementation of advertising networks on websites and services, such as GoogleAds\citep{TAHAEI2021}.

\subsection{Connecting the Research Questions - Discussion}
\label{sec:mappingresultsdiscussion}
The diversity of deceptive pattern types identified in RQ1 is mirrored in the heterogeneous terminology and taxonomies discussed in RQ3, pointing to a fragmented field with overlapping and sometimes conflicting definitions. This conceptual fragmentation corresponds to the scarcity of theoretical grounding observed in RQ5, where most works rely on ad hoc notions rather than established theoretical perspectives. Such limited theorization, in turn, contributes to the methodological fragmentation identified in RQ4, as many empirical investigations lack formal frameworks or standardized procedures. The dominance of user-centered approaches found in RQ4 further explains why RQ6 reveals a strong emphasis on human and ethical dimensions, while technical and infrastructural aspects remain underexplored. 

Finally, the concentration of studies in the upper layers of the Semiotic Framework (RQ7) and the limited attention given to the technical and physical layers (RQ8) confirm this imbalance: deceptive patterns are predominantly treated as ethical issues rather than as truly sociotechnical constructs. This imbalance reflects how the current scientific discourse itself feels a lot similar to capitalist dynamics: while the market incentivizes research on user behavior and engagement, less emphasis is placed on the technical mechanisms that enable such manipulation. These findings form the empirical foundation for the theoretical discussion that follows, where we interpret deceptive patterns as expressions of socio-economic forces embedded in digital capitalism.

Taken together, RQ1-RQ8 depict a research field characterized by conceptual dispersion, theoretical fragility, and methodological heterogeneity. The emphasis on human, ethical, and behavioral aspects contrasts with the limited attention to technical and infrastructural dimensions. This pattern reveals the need for a interconnected, considering both social impacts and technical details, sociotechnical understanding of deceptive design - an analytical gap addressed in the following section through the application of the Semiotic Framework.

\section{Goal 3. Sociotechnical Aspects of the Mapped Literature}
\label{sec:goal3}

To achieve the third specific objective-analyzing the sociotechnical aspects considered and overlooked in the literature-this section presents the results of the systematic mapping under the light of the Semiotic Framework (Section \ref{sec:theframework}), applied to examine the selected articles from a systemic perspective across its six distinct layers, and the results obtained.

\begin{figure}[h]
    \centering
    \includegraphics[width=0.48\textwidth]{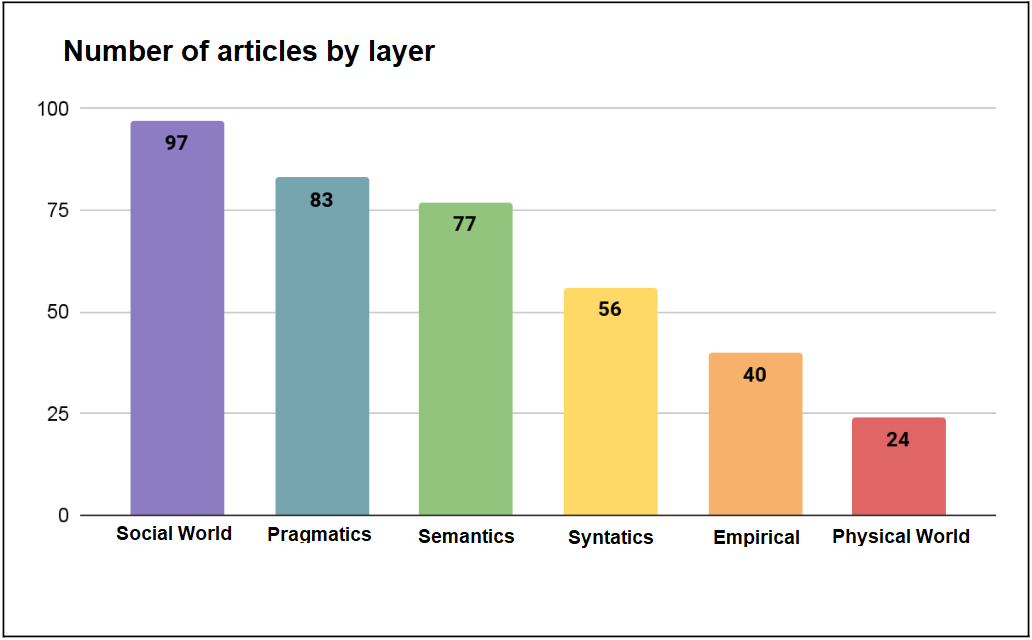}
    \caption{Number of articles classified within each layer.}
    \label{fig:layers}
\end{figure}

Articles were analyzed and categorized according to the Semiotic Framework layers, as depicted in Figure \ref{fig:layers}. Many articles fit into multiple layers. A key observation is that most studies emphasize the human-functional layers (social world, pragmatic, and semantic), focusing on social impacts, user experiences, expectations, meanings, ethical considerations in interface design, human values in interface creation, legal frameworks, user behavior, and manipulation strategies. In contrast, fewer studies address the technical-platform layers (syntactic, empirical, and physical world), such as visual, structural, and statistical characteristics of deceptive interfaces, their material and temporal costs to users, and how the physical environment is leveraged to influence users.

All selected articles in the \textbf{Social World Layer} discuss the societal and user-level impacts of deceptive interfaces, including privacy violations, security threats, behavioral and emotional manipulation, and consent exploitation. Many refer to Value Sensitive Design \citep{FRIEDMAN2008,FRIEDMAN2019} as a primary approach to mitigating these adverse design effects. Several works present user perspectives, experiences, and emotional responses to deceptive patterns.

Additionally, some studies examine the relationship between deceptive patterns and legal frameworks, including the European GDPR \citep{GDPR}, California’s CCPA \citep{CCPA}, and Brazil’s LGPD \citep{LGPD}. These regulations define free and informed consent, specify data collection and usage limits, and require mechanisms for user control and data deletion. Many articles focus on cookie consent banners that employ deceptive patterns to maximize user agreement, thus violating these laws. Despite jurisdictional differences, these laws share common principles: interfaces must provide straightforward options for data deletion and consent refusal, ensuring informed user choices. The literature within this layer offers a comprehensive overview of deceptive patterns' social, legal, emotional, and behavioral impacts.

In the \textbf{Pragmatic Layer}, many studies highlight the conflict between user intentions and deceptive interface objectives. Grounded in psychology, behavioral economics, or neuromarketing, these studies explain how deceptive patterns deliberately exploit cognitive biases \citep{MATHUR2019} and employ A\/B testing \citep{NARAYANAN2020,ILANYTZUR2016} to optimize manipulation strategies.

Article \citep{NARAYANAN2020} identifies three key influences on the emergence of deceptive patterns: long-standing deceptive retail practices, HCI research on nudges (light manipulation to enhance usability), and growth hacking (strategies to accelerate business growth). Only two studies \citep{MA2022,CHIVUKULA2019b} explicitly examine interface designers' intentions through simulated experiments on deceptive interface design.

In the \textbf{Semantic Layer}, numerous studies analyze deceptive patterns and conduct user experiments, revealing the true deceptive nature of these interfaces. Users often perceive them as neutral or honest, despite their deceptive design. Most studies addressing the Syntactic Layer also contribute to the Semantic Layer.

For instance, \cite{TAHAEI2021} evaluates developer guides for online ad platforms like Google Ads. The study reveals that companies shift privacy responsibilities onto developers, contradicting the perspective of both developers and researchers, who argue that ad platform providers should bear accountability. This behavior of companies suggests an intentional strategy by companies to evade responsibility for user privacy.

In the \textbf{Syntactic Layer}, many articles also contribute to the Semantic Layer. Eleven studies demonstrate, through user experiments, how the formal structures of deceptive interfaces manipulate behavior. These studies assess different consent banner designs, button placements, colors, and formats, showing how specific layouts significantly increase consent rates at the expense of user autonomy. Two articles specifically analyze the language used in online contracts, examining formality levels and information granularity to assess their role in influencing user decisions.

For example, \cite{PETROVSKAYA2022} demonstrates that the wording in cookie consent prompts can substantially impact user decisions. Another study \citep{LUKOFF2021a} analyzes YouTube’s mechanisms that diminish user control over content consumption. Additionally, some articles examine deceptive interface patterns in websites, apps \citep{KOLLNIG2021,GUNAWAN2021,CHEN2022}, and games \citep{DAS2022,FITTON2019}, identifying formal deceptive structures.

In the \textbf{Empirical Layer}, fewer studies explore the statistical properties and platform capabilities that deceptive patterns exploit, such as long and complex legal terms, varying constraints across platforms (e.g., Facebook, Spotify, Wish), or misleading cookie banners. In \citep{MOSER2019}, the authors discuss interface features that encourage impulsive purchases. In \citep{DICKINSON2022}, a study speculates on future risks posed by human-machine integration, including parasitic symbiosis, where machines exploit user emotions and could even override bodily functions.

In the \textbf{Physical World Layer}, a limited number of articles examine deceptive patterns' material and economic consequences. Some studies discuss Surveillance Capitalism, highlighting how this economic model normalizes unethical mass surveillance and user manipulation.


The mapped literature predominantly focuses on the social aspects of deceptive patterns, including their negative impacts (social world), user intentions (pragmatic), and the deceptive meanings embedded in interfaces (semantic). However, many analyses remain narrowly focused on specific subtopics, such as categorizing deceptive patterns, presenting user opinions, or discussing ethical considerations. Less explored areas include the origins and funding of these patterns, corporate exploitation of transmission media to manipulate users, and how deceptive patterns leverage the user's physical environment.

Further research could examine developers' roles and motivations in implementing deceptive designs within the Pragmatic Layer. Are these patterns solely imposed by higher management? How can developers counteract this growing culture of manipulation within Surveillance Capitalism? There is also a significant opportunity to investigate the technical characteristics of deceptive patterns, particularly their structural composition.

\section{Goal 4. The Self-Explanatory Online Catalog}
\label{sec:goal4}
To achieve the fourth specific objective-develop a catalog of the mapped deceptive patterns-the online self-explained catalog of deceptive patterns was built after the systematic mapping. The structure of the website was built using the NicePage \footnotetext{https://nicepage.com/. Accessed on February 21, 2026} website builder for easier construction and faster deployment than programming and building the website from the ground-up. 

All the deceptive patterns found during the systematic mapping were added in the catalog under 11 different categories. These categories resulted from the systematic mapping through an incremental emergent exploratory coding inspired by Thematic Analysis \citep{braun2012thematic}. The different types of patterns found were compared pairwise, and based on the original categories of \cite{BRIGNULL2023}, \cite{MATHUR2019}, and \cite{GRAY2018}, new categories were created or existing ones merged until 11 main categories remained. These categories were presented in Section \ref{sec:mappingresults}. 

\subsection{The Deceptive Patterns Online Catalog}
\label{sec:catalog}

Organizing and structuring the mapped deceptive patterns, a catalog was created containing all the deceptive patterns identified in the systematic mapping. This catalog is available in Portuguese online at the following URL\footnote{https://deceptivepatterns.nicepage.io/}.

\begin{figure}[h]
\centering
\includegraphics[width=0.48\textwidth]{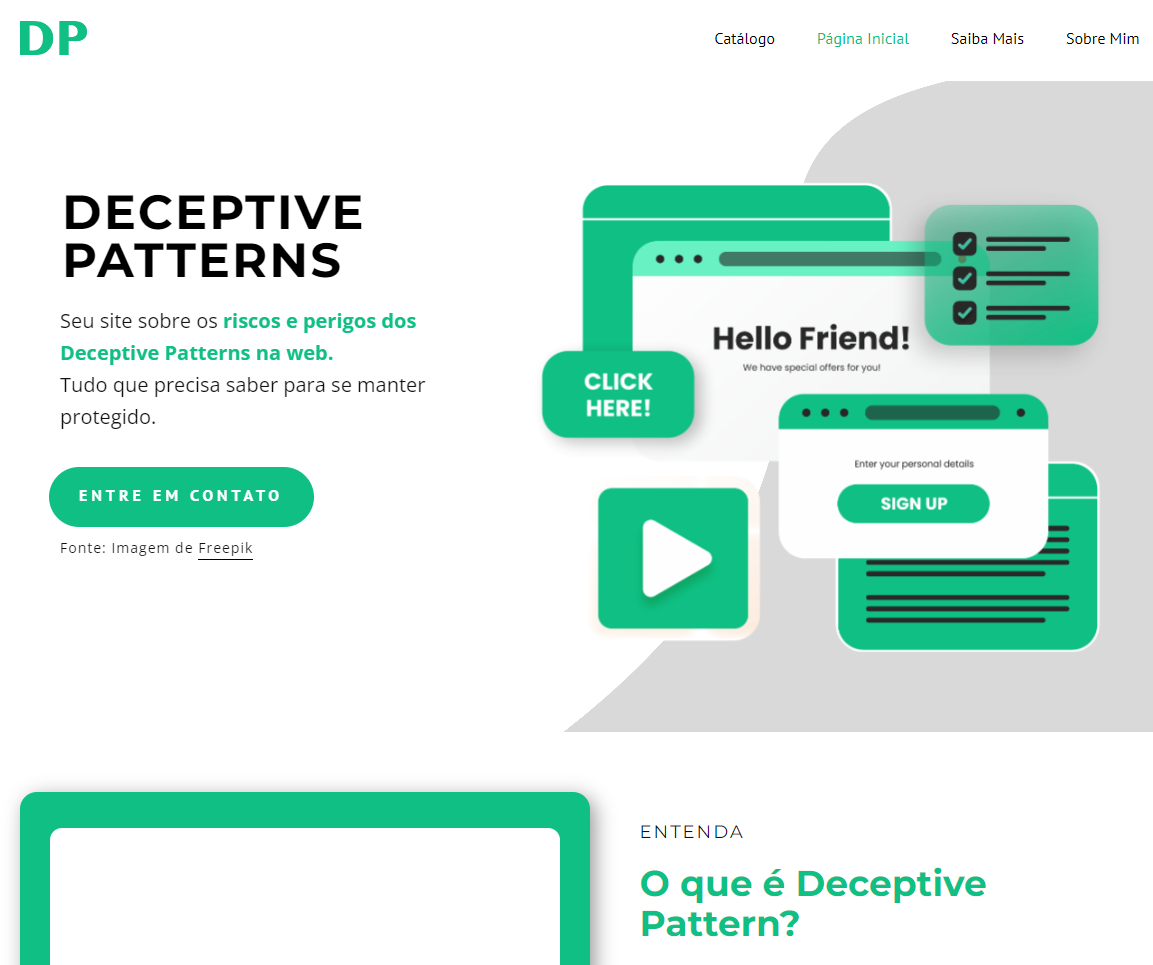}
\caption{Homepage of the website.}
\label{fig:paginicial}
\end{figure}

The catalog is a technical contribution designed to present information on deceptive patterns documented in the literature. It serves as a resource for design practices and education, addressing this sociotechnical phenomenon, which remains largely unknown outside academic circles. By organizing and explaining how deceptive patterns function and their effects, the catalog helps users recognize these unethical practices, enabling them to avoid and protect themselves. Additionally, it serves as a resource for professionals seeking to understand and prevent these deceptive design strategies. 

All the 126 deceptive patterns found during the systematic mapping are listed in the catalog. Each one of the 11 categories has a page in the catalog, where all the patterns under that category are listed alongside a brief explanation.

\begin{figure}[h]
\includegraphics[width=0.48\textwidth]{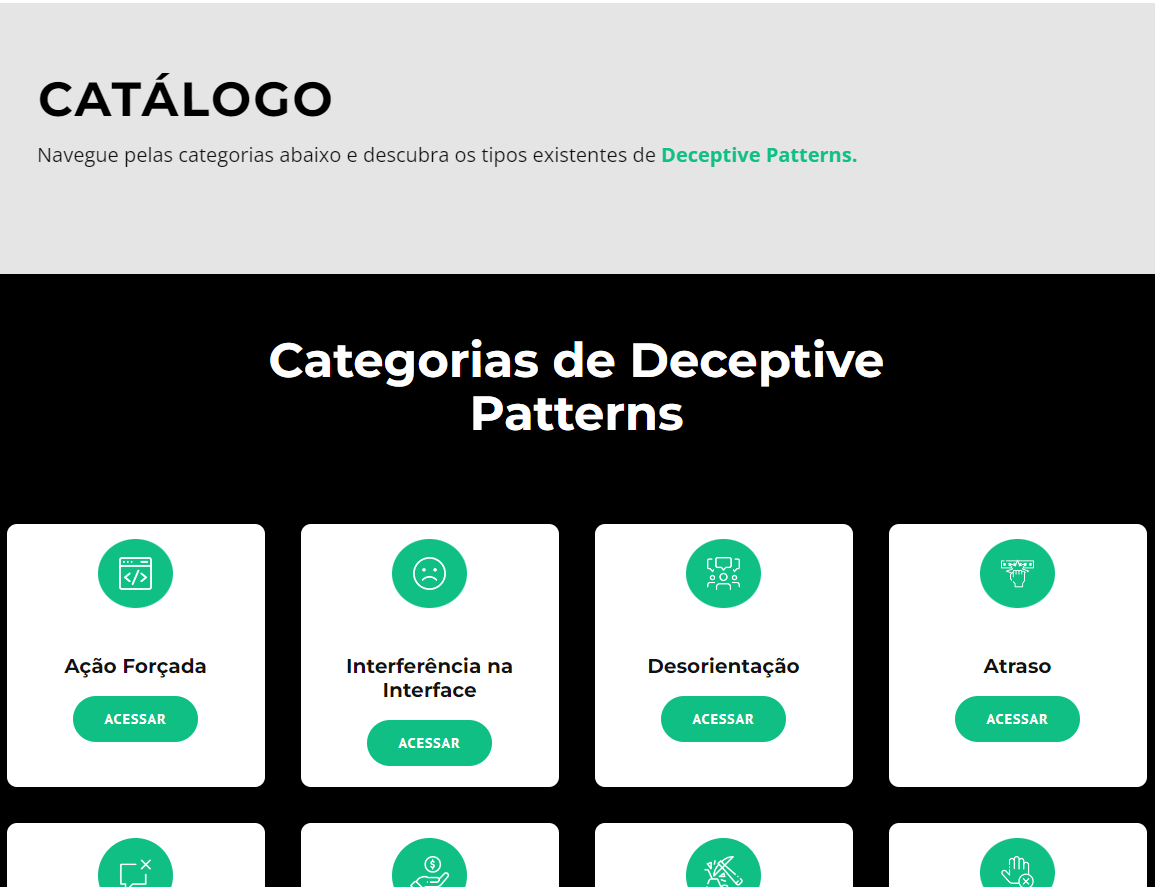}
\caption{Catalog page displaying some of the categories.}
\label{fig:catalogo}
\end{figure}

\begin{figure}[h]
\includegraphics[width=0.48\textwidth]{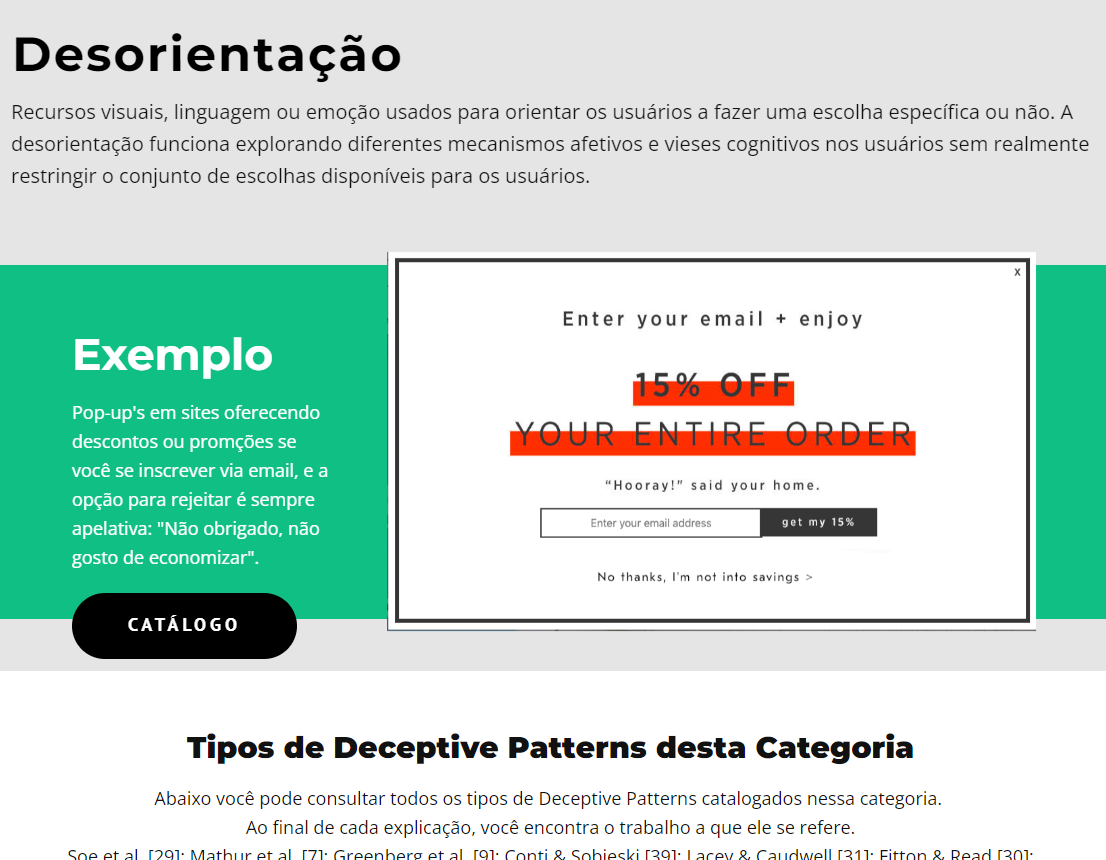}
\caption{Screen showing the "Disorientation" category.}
\label{fig:acaoforcada}
\end{figure}

\subsubsection{Catalog Homepage}

The first section of the homepage (Figure \ref{fig:paginicial}) includes a top-right menu with four options: Catalog (leading to the deceptive patterns catalog), Homepage (returning to the main page), Learn More (providing additional information about deceptive patterns and their mechanisms), and About Me (offering details about the author of this work). Below the menu is a large image on the right illustrating a deceptive pattern. In contrast, on the left, the title "Deceptive Patterns" is displayed alongside an introductory phrase: "Your site for understanding the risks and dangers of deceptive patterns on the web. Everything you need to stay protected." Beneath this text is a "Contact Us" button.

\begin{figure}[h]
\includegraphics[width=0.48\textwidth]{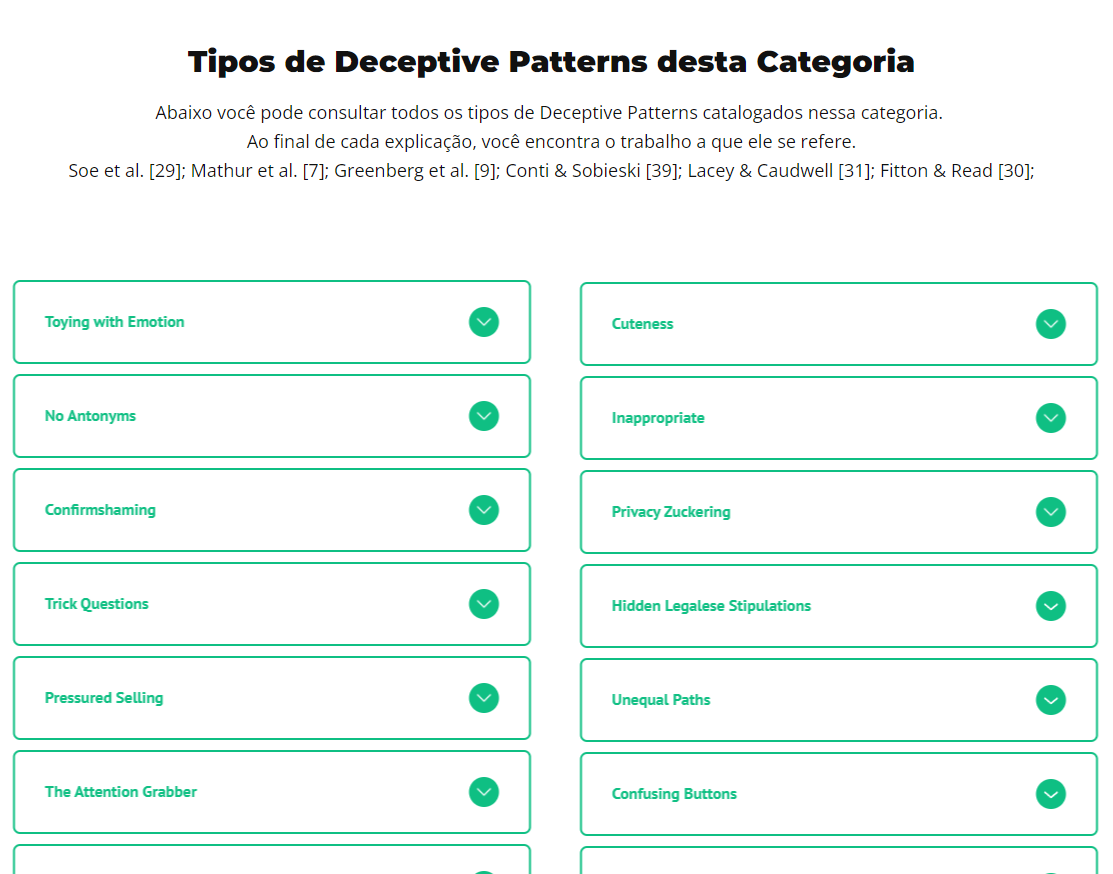}
\caption{Different types of deceptive patterns within the "Disorientation" category.}
\label{fig:tiposacao}
\end{figure}

\begin{figure}[h]
\includegraphics[width=0.48\textwidth]{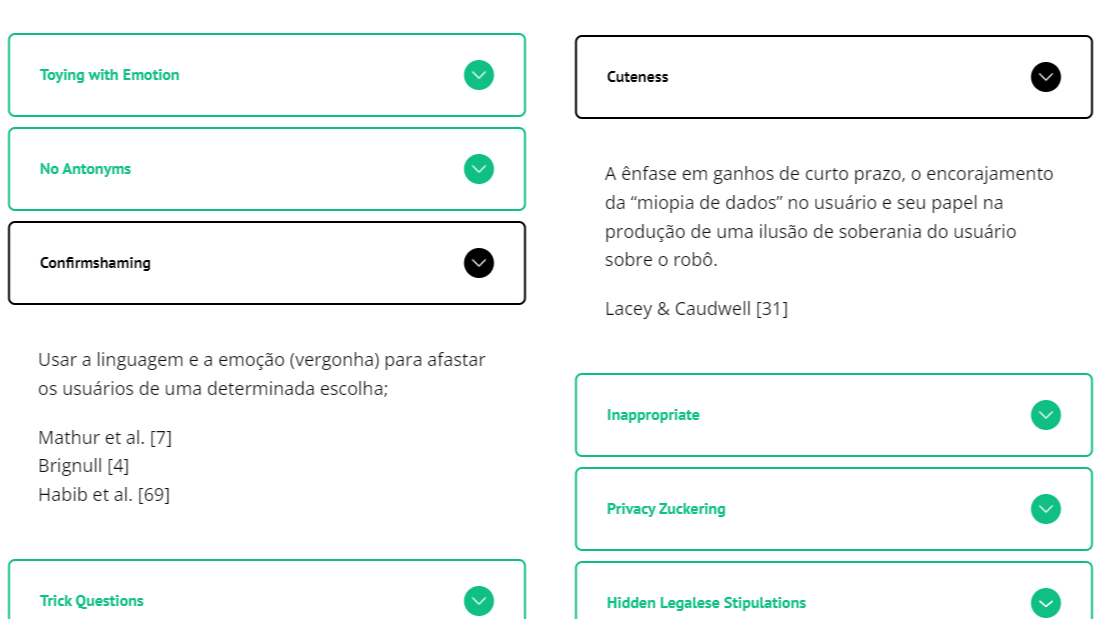}
\caption{"Cuteness" and "Confirmshaming" types.}
\label{fig:categorias}
\end{figure}

\subsubsection{Catalog Page}

The catalog page (Figure \ref{fig:catalogo}) features a prominent title at the top, followed by an introductory phrase: "Browse the categories below and discover the existing types of deceptive patterns." Below this, the eleven deceptive pattern categories are presented in separate boxes, each containing a button that navigates to the corresponding category page.

\begin{figure}[h]
\centering
\includegraphics[width=0.48\textwidth]{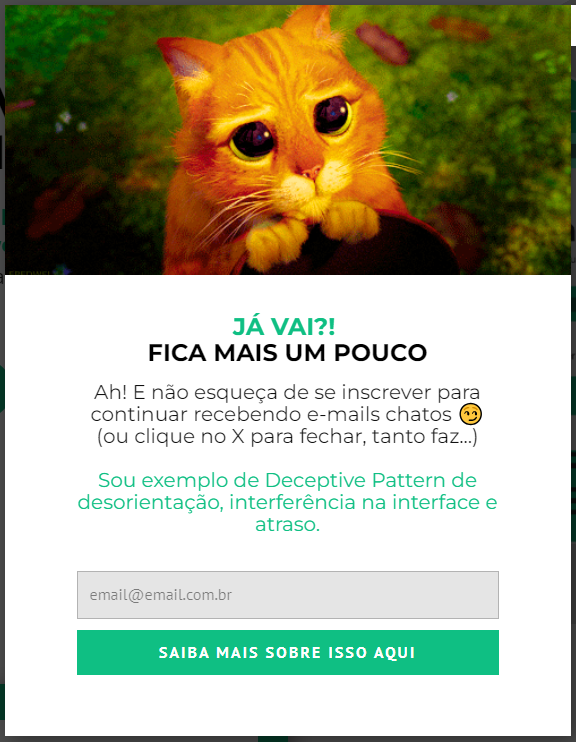}
\caption{Example of disorientation.}
\label{fig:javai}
\end{figure}

\begin{figure}[h]
\centering
\includegraphics[width=0.48\textwidth]{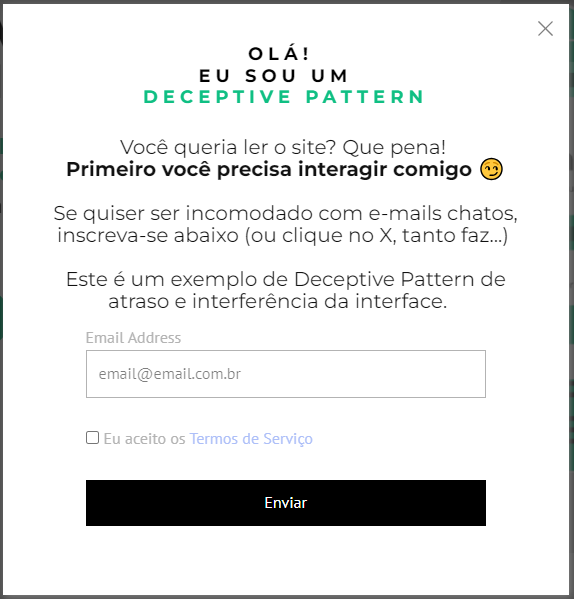}
\caption{Example of obstruction.}
\label{fig:quepena}
\end{figure}

When users select a category, they are directed to a page displaying the category's title, a description, a practical example, and, further down, a two-column layout listing all deceptive pattern types that fall under this category. Figure \ref{fig:acaoforcada} captures the "Disorientation" category screen, showing an example and a brief explanation, while Figure \ref{fig:tiposacao} presents different types within this category.

Clicking on a specific type reveals its description. Figure \ref{fig:categorias} provides explanations for the "Confirmshaming" and "Cuteness" types: 
\begin{itemize}
    \item \textbf{Confirmshaming}: Using language and emotions (such as shame) to deter users from making a particular choice.

    \item \textbf{Cuteness}: Exploiting short-term gains, encouraging users’ "data myopia," and fostering an illusion of control over domestic robots.
\end{itemize}

To enhance its self-explanatory nature, a subset of deceptive patterns was implemented to appear during user navigation, offering practical examples that mimic deceptive patterns and provide an educational, hands-on experience. For instance, on the homepage, a pop-up appears a few seconds after loading or when the users moves the cursor away from the page, blocking interaction and requesting the user’s email (Figures \ref{fig:javai},\ref{fig:quepena}). To the best of our knowledge, this is the only catalog derived from a systematic literature mapping and the first of its kind in Portuguese. Additionally, no other tool has been developed to raise awareness of deceptive patterns and inform users about their characteristics and mechanics during interaction.

\subsection{Evaluation of the Catalog}
\label{sec:catalogEv}

In the final stage, the catalog was evaluated for its Ease of Use and Perceived Usefulness by seven participants with previous experience with HCI design and evaluation techniques. The evaluation was done in three parts: at first an initial phase of free exploration of the website by participants, then a questionnaire based on the Technology Acceptance Model (TAM) \citep{Davis1989}, using a 5-point Likert scale (1 - Strongly Disagree; 2 - Disagree; 3 - Neutral; 4 - Agree; 5 - Strongly Agree) to evaluate usability and perceived usefulness, and by the end a focus group with the participants. The questionnaire consisted of participant characterization, usability assessment, and perceived usefulness evaluation. The questionnaries used can be accessed here: \href{https://docs.google.com/spreadsheets/d/178Ce3HrhzRiUocvI5JiAhsjFdM0hYrP04kn6pVNW8_0/edit?usp=sharing}{\textcolor{blue}{\textbf{\underline{Questionnaries}}}}. 

The evaluation process occurred as following: Participants first read and signed the informed consent form. Then participants navigated the Deceptive Patterns catalog website for around 15 minutes and explored its content. During this phase, participants were instructed to access the “Catalog” section, navigate through different categories and types of deceptive patterns, and the “Learn More” section. Then they were let free to continue interacting with the website as they wished until the 15 minutes ended. After exploring the website, participants answered the digital questionnaire, taking up to 30 minutes for completion. 

After the questionnaire, all participants joined in a focus group to further elaborate on their perceptions of the catalog from the previous study. The interaction dynamic followed an open discussion format: the author informally asked participants about their thoughts on the study and the catalog. Each participant then verbally contributed to the discussion regarding the acceptance of the online catalog. The conversation was recorded via audio, later transcribed, and analyzed.
\subsubsection{Ethical Considerations}
\label{ssec:ethical}

Computational techniques are not neutral \citep{carvalho2021ethics}, and this research carries ethical implications by examining user interface patterns that may manipulate behavior and decision-making. This study highlights concerns about user privacy, consent, and deceptive design practices, emphasizing the need to identify and catalog deceptive patterns to mitigate harm and promote ethical design.

An ethical issue encountered was the terminology used. Initially, we adopted Brignull's term "\textit{dark patterns}" \citep{BRIGNULL2023}, but later discussions indicated it could be offensive or exclusionary\footnote{See ACM Words Matter page: https://www.acm.org/diversity-inclusion/words-matter. Accessed on February 22, 2026.}. Given its non-essential nature, we adopted "deceptive patterns," a more neutral and self-explanatory term.

Regarding the ethical considerations associated with participant involvement in the exploratory catalog evaluation, the study was conducted exclusively with members of our research laboratory. Such studies are necessary before studies involving other audiences. Participation was voluntary, no personal or sensitive data were collected, and all activities performed fall within the scope of the laboratory’s routine academic practices. Accordingly, we consider that this exploratory evaluation falls under Article 1 of Resolution No. 510/2016 of the National Health Council (Conselho Nacional de Saúde, Brazil), which states that research aimed at the theoretical deepening of situations that emerge spontaneously and contingently in professional practice, and that does not disclose data capable of identifying individuals, does not require registration with or evaluation by the CEP/CONEP system.\footnote{See \href{https://bvsms.saude.gov.br/bvs/saudelegis/cns/2016/res0510_07_04_2016.html}{Ministry of Health page}. Accessed on February 22, 2026} Nevertheless, we acknowledge the need for further evaluation activities involving broader audiences, and such studies are currently underway. For example, in a separate research project approved by an Ethics Committee \citep{verhalen2025deceptive}, the catalog was explored by older adult participants. The results indicated increased awareness and recognition of deceptive patterns; however, barriers persisted, including the use of technical language, aspects of the catalog’s interface design, and participants’ limited familiarity with the topic.



\subsection{Results of Evaluation}

An exploratory evaluation with HCI professionals -- colleagues from the research laboratory -- was conducted to obtain first insights about the usability and perceived usefulness of the catalog as well as to anticipate problems that must be fixed before evaluating the catalog with other audiences. Seven participants explored the catalog and answered two questionnaires: first, a characterization questionnaire, and second a questionnaire based on the Technology Acceptance Model. In the following, participants discussed the results and their impressions in a focus group. Table \ref{tab:catalog_study} provides further details on the first part of the study.

\begin{table}[]
\centering
\begin{tabular}{|l|l|}
\hline
\textbf{Objective} &\begin{tabular}[c]{@{}l@{}}To conduct a preliminary evaluation \\of the online catalog of deceptive \\ patterns through an analysis of \\ its usability and perceived \\ usefulness by HCI professionals.\end{tabular} \\\hline
\textbf{Target Audience} & \begin{tabular}[c]{@{}l@{}}Computer science researchers, \\ and professionals in interface \\design and development.\end{tabular} 
\\\hline
\textbf{Total Duration} &\begin{tabular}[c]{@{}l@{}}1 hour and 5 minutes (45 minutes \\ for the questionnaire activity and 20\\ minutes for the focus group).\end{tabular}  
\\\hline
\textbf{Location} &  \begin{tabular}[c]{@{}l@{}}Laboratory 3, Department of\\ Informatics, Federal University\\ of Paraná, Curitiba, Brazil.\end{tabular} 
\\\hline
\textbf{Date} & \begin{tabular}[c]{@{}l@{}}September 29, 2022.\end{tabular} 
\\\hline
\textbf{Participants} & \begin{tabular}[c]{@{}l@{}}Seven (7) professionals with \\experience in HCI.\end{tabular}
\\ \hline
\end{tabular}
\caption{Summary of the study conducted with the catalog.}
\label{tab:catalog_study}
\end{table}

The results from the characterization questionnaire indicates that the study participants represent part of the target audience for the deceptive patterns catalog. All seven participants reported using the internet daily, making them susceptible to encountering and being affected by deceptive patterns. Furthermore, all participants have a background in Computer Science and work as HCI researchers.

Questions 4 and 5 of the characterization questionnaire focused on participants' knowledge of deceptive patterns before and after using the catalog. Six out of seven participants reported increased knowledge of deceptive patterns after utilizing the catalog.

\subsubsection{Usability and Perceived Usefulness}

\begin{figure}[h]
    \centering
    \includegraphics[width=0.48\textwidth]{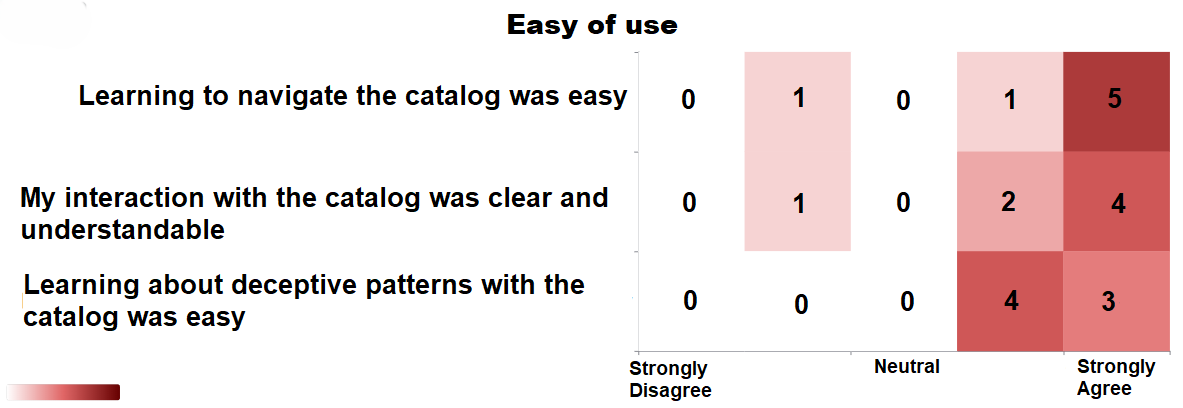}
    \caption{Usability questionnaire.}
    \label{fig:usabilityHM}
\end{figure}

\begin{figure}[h]
    \centering
    \includegraphics[width=0.48\textwidth]{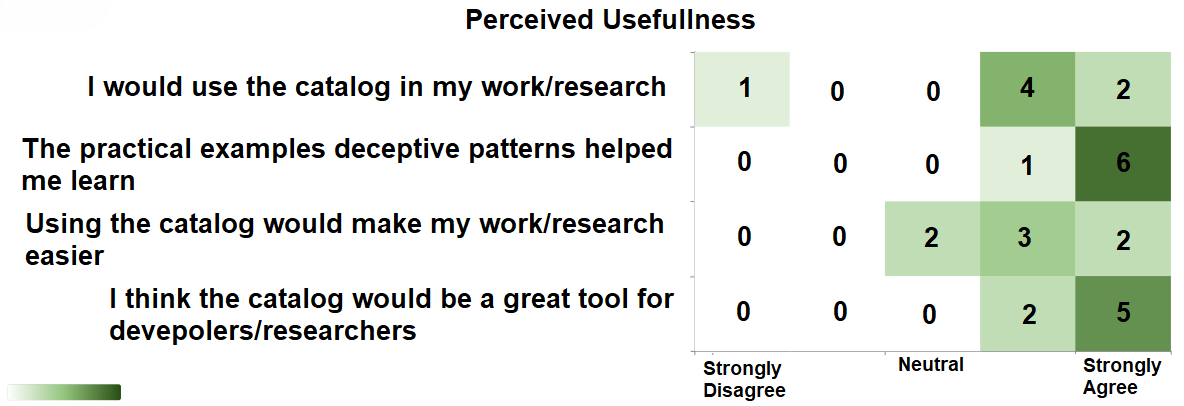}
    \caption{Perceived Usefullness questionnarie.}
    \label{fig:percieveduseHM}
\end{figure}

Following the characterization questionnaire, participants answered usability and perceived usefulness questionnaires for the catalog using a 5-point Likert scale: 1-Strongly Disagree, 2-Partially Disagree, 3-Neutral, 4-Partially Agree, 5-Strongly Agree. Figures \ref{fig:usabilityHM} and \ref{fig:percieveduseHM} present the questions and participants' responses.


Regarding usability (Figure \ref{fig:usabilityHM}), the study results indicate that the catalog is easy to use and learn. Six participants agreed with the statements “It is easy to learn how to navigate the catalog” and “The interaction with the catalog was clear and understandable.” All participants agreed that learning about deceptive patterns using the catalog was easy. Three participants left comments reinforcing this ease of use:
\begin{itemize} 
\item Participant 2 - “The catalog is great, very easy to use.”
\item Participant 6 - “... easy to interact with due to the minimalist design.”
\item Participant 7 - “I found the catalog easy to navigate.”
\end{itemize}

All participants stated that the practical examples of deceptive patterns implemented on the catalog website helped them understand them. Five participants agreed using the catalog would make their work/research easier. All participants believed the catalog would be a valuable tool for developers and researchers. Six participants stated that they would use the catalog in their work/research. The combined results highlight the catalog's ease of use and perceived usefulness. in teaching deceptive patterns to laypeople, including users, researchers, and developers (Figure \ref{fig:percieveduseHM}).

Some participant comments confirming the catalog's usefulness:
\begin{itemize}
\item Participant 3 - "Perhaps the deceptive patterns catalog is most important for less experienced users, serving as an educational tool on deceptive patterns."
\item Participant 4 - "The catalog seems to be highly useful."
\item Participant 6 - "The website is very effective in providing a didactic, engaging, and informative way to understand deceptive patterns."
\item Participant 7 - "I can see myself using this deceptive patterns catalog when developing systems or conducting research on the topic. I also believe the catalog could be beneficial in an educational setting. I can imagine it being used in discussions on interface design or even in an HCI class."
\end{itemize}

\subsubsection{Focus Group}

The focus group was important to further elaborated on the perceptions of participants about the catalog, enabling a more refined final product.



During the focus group participants expressed enthusiasm and interest in the catalog, offering positive feedback and suggestions for improving the website and catalog. Some participants were particularly excited about the practical examples of deceptive patterns implemented on the catalog website:

\begin{itemize}
\item "... browsing through the categories and reading about the different deceptive patterns, I felt it was a very interesting way to explore them..." "...getting an overview of what each one is, I found it an engaging way to discover and learn about them..."
\item "I thought the deceptive patterns presented on the website were excellent..."
\item "...what stands out on the site, beyond the knowledge itself, is how it interactively demonstrates what a deceptive pattern is..." "...in terms of usability, those elements shouldn't be there, but from a user experience perspective, it's phenomenal-super fun to interact with these features..."
\end{itemize}

Participants also provided various suggestions for improving the website and catalog, including:
\begin{itemize}
\item Animated GIFs for all deceptive pattern examples;
\item A quiz on deceptive patterns to test catalog visitors;
\item Experimental educational games about deceptive patterns;
\item A questionnaire to help users identify which deceptive pattern is affecting them;
\item Ideas on how to monetize the catalog;
\item And other suggestions.
\end{itemize}

Participants also pointed out typos and website implementation issues they encountered during the study, which were addressed and corrected.

\section{Strategy of Capitalism}
\label{sec:capitalism}

The systematic mapping and sociotechnical analysis showed that deceptive patterns are not isolated interface features but manifestations of structural incentives embedded in digital economies. From the results of the Systematic Mapping, (Section \ref{sec:mappingresults}) the fact that most studies frame deception through ethical or behavioral lenses, with little focus on the underlying technical infrastructures, indicates that manipulation is often naturalized as part of design practice rather than questioned as a systemic phenomenon. Building upon these empirical insights, we now discuss how the capitalist logic of optimization, efficiency, and profit-seeking sustains and normalizes deceptive practices in design.

Ultimately, deceptive patterns are another strategy of capitalism, a result of the prevailing production model in our society. They arise from the overvaluation of profit at all costs, at the expense of human beings and the impacts this pursuit can cause. The relentless search for ways to profit from everything that permeates our lives has led companies and businesses to create increasingly efficient methods of capturing people's minds and desires, making them hostages to their deepest instincts or fears. In other words, deceptive patterns emerge when companies and businesses ignore that their customers are, at their core, complex human beings with diverse realities and a lifetime of needs, desires, responsibilities, etc. Deceptive patterns arise when humans are treated merely as "consumers" or "users" \citep{GONZATTO2022}. This is the result of the commodification of life in society.

A deleterious practice that only harms society and brings little or no benefit to those who use it, as, from a psychological perspective, when a person receives negative reinforcement (being deceived or harmed in any way, whether monetarily or temporally), they tend to avoid the situation that left them with a bad feeling. Therefore, deceptive patterns are not only harmful for those who suffer from them but, in the long term, are also detrimental to those who use them.

Although there is no single study confirming that deception will continue indefinitely, the combination of research in psychology, history, and design suggests that as long as there are incentives to manipulate human behavior, new forms of deceptive patterns will continue to emerge and evolve. One solution that has been effective against deceptive patterns in other countries is legislation, such as the GDPR and the CCPA, which have driven fundamental changes in online cookie banners. There is a need for laws explicitly addressing the predatory behavior of interfaces and companies in virtual/digital spaces, as the European GDPR and the Californian CCPA do. Through regulation, oversight, and control of these practices, as is already done with many other practices commonly known to violate consumer rights, it will be possible to reduce the impacts of deceptive patterns on society and punish those who use them.

While deceptive patterns originate from corporate strategies, it is essential to recognize that their implementation depends on developers-those who design and build digital experiences. Developers, however, often find themselves in a paradoxical position: they are both enforcers of deceptive practices and, at times, victims of the corporate environment that demands their execution. Many work under strict directives, constrained by business goals and performance metrics that prioritize engagement and revenue over ethical design. In such environments, refusing to implement deceptive tactics can result in professional repercussions, from being sidelined in projects to job insecurity. Thus, while developers are responsible for translating corporate strategies into code, their agency is often limited by the very system that commodifies users. Addressing deceptive patterns, therefore, requires not only regulating corporations but also advocating for ethical work environments where developers are empowered to prioritize user well-being over profit-driven coercion.

Finally, it is also crucial to focus on raising awareness about these practices. Knowledge is an excellent way to empower people so that they can, on their own, make decisions that benefit their lives and avoid falling victim to scams and misleading advertisements.

By linking the mapping results to the socio-economic interpretation of deceptive design, this study demonstrates how deceptive patterns are sustained not only by individual design decisions but by the structural logic of capitalist production. The Semiotic Framework made it possible to reveal this interplay between social norms, technological affordances, and economic motivations -reinforcing the need for sociotechnical approaches to design ethics.

\section{Conclusion}
\label{sec:conclusion}


This work addressed the issue that the literature presents discussions and analyses focused on and limited to specific aspects of deceptive patterns, such as merely listing and categorizing deceptive patterns, presenting user opinions on these interfaces, or discussing societal themes like ethics and values related to deceptive patterns. Consequently, other fundamental aspects of deceptive patterns are often overlooked, such as who created/funded these patterns, how companies exploit the transmission medium to hinder and manipulate, and how these patterns can even take advantage of the user's physical environment.

Thus, the objective of this work was: \textbf{"To characterize deceptive patterns as a sociotechnical phenomenon in Computing."} To achieve this goal, the following specific objectives were fulfilled:
\begin{enumerate}
    \item Analyze and characterize a deceptive pattern from a sociotechnical perspective;
    \item Map the literature on deceptive patterns in computing;
    \item Analyze the sociotechnical aspects considered and neglected in the literature;
    \item Develop a catalog of the mapped deceptive patterns.
\end{enumerate}

To satisfactorily fulfill these objectives, the methodology employed in this work was based on an analytical and constructive investigation. The \textbf{analytical investigation} was conducted in three stages:
\begin{enumerate}
    \item Stage 1: An analysis of a deceptive pattern grounded in the Semiotic Framework [30];
    \item Stage 2: A systematic mapping of the literature on deceptive interfaces in the main indexing databases for computing articles (ACM, IEEE, and Scopus);
    \item Stage 3: An analysis of the sociotechnical aspects considered and neglected in the literature, also based on the Semiotic Framework.
\end{enumerate}

In the \textbf{constructive investigation}, as a technical contribution, a catalog containing the deceptive patterns identified in the scientific literature was organized, constructed, and evaluated by a group of seven professionals with experience in HCI.

Fulfilling the first objective (Analyze and characterize a deceptive pattern from a sociotechnical perspective), in Section \ref{sec:goal1} the analysis and characterization of a deceptive pattern as a sociotechnical phenomenon is demonstrated through the analysis of the "Roach Motel" deceptive pattern. It is shown how deceptive patterns exploit elements of the interface, language, and even human psychology. These elements range from the user's social world to the physical medium through which the users interacts with the deceptive pattern in a systemic and interconnected manner.

Fulfilling the second objective (Map the literature on deceptive patterns in computing;), in Section \ref{sec:goal2}, the mapping of the literature on deceptive patterns provided a rigorous overview of the field of deceptive patterns in Computer Science. Despite the extensive literature on deceptive patterns, most of it has been produced in the last five years (from 2018 to the present). However, most authors fail to consider deceptive patterns systematically or as a sociotechnical phenomenon. There are also exhaustive discussions and analyses around cookie warnings and how they function and affect users.

Fulfilling the third objective (Analyze the sociotechnical aspects considered and neglected in the literature;), in Section \ref{sec:goal3}, after analyzing the sociotechnical aspects considered and neglected by the literature, it was discovered that most articles favor the social aspects of deceptive patterns (those in the Social World, Pragmatic, and Semantic layers) and disregard the technical aspects of deceptive interfaces (in the Syntactic, Empirical, and Physical World layers), which are fundamental to explaining how and why they exist and function.

Finally, fulfilling the fourth objective (Develop a catalog of the mapped deceptive patterns.), in Section \ref{sec:goal4}, the catalog containing all the deceptive patterns found in the literature was constructed and evaluated for its ease of use and perceived usefulness. The study involved seven participants who also participated in a focus group after the study to understand their opinions better. Regarding ease of use, participants indicated that the catalog is easy to use and learn and that learning about deceptive patterns through the catalog is also straightforward. Regarding perceived usefulness, participants suggested that the catalog is an excellent tool for developers and researchers. It was also found that using the catalog would make the participants' work or research easier and that they would use the catalog in their work or study.

According to the study participants, the practical examples of deceptive patterns implemented in the catalog were helpful in teaching about deceptive patterns. All participants agreed with the statement, "The practical examples of deceptive patterns in the catalog helped me learn about the topic." After the study, one participant stated that they learned more from the practical examples than the rest of the catalog. Showing the didactic potential of practical examples of deceptive patterns, enabling awareness of this topic within a controlled environment by placing users in "real" situations without the associated risks.

The results of this work, taken together, indicate that deceptive patterns are a dangerous phenomenon for the well-being of all internet users, potentially causing negative social, emotional, financial, and temporal impacts. These patterns act in favor of companies, businesses, online content creators, programmers, and designers, always at the user's expense, often in ways that users do not even notice. This phenomenon has been occurring for a long time across various user interfaces, from mobile phones to computers to self-service kiosks, in digital and print media, and even in physical spaces. However, it has only gained the attention it deserves in the last six years. Therefore, this work actively contributes to advancing the understanding of deceptive patterns and helps researchers, users, and developers. The results allow us to view deceptive patterns as a phenomenon that urgently needs to be addressed.

\subsection{Limitations and Threats to Validity}

The main challenges and limitations of this work were: the language used in the mapping was limited to English, which may have excluded important articles in Portuguese; during the mapping, the search string was incomplete, omitting the term \textit{manipulative}, which could have added more articles of interest to the research; the sociotechnical analysis was conducted with only one specific pattern, which may lead to a partial understanding of the whole.

There is a lack of prior research that has analyzed deceptive patterns using the Semiotic Framework, which could have supported this study; the catalog acceptance study was conducted only with HCI professionals, limiting the possible conclusions, and a broader follow-up study with the general public is necessary for confirmation.

\subsection{Contributions and Future Works}

As outcomes of this research, this work contributes to science by advancing the community's knowledge of deceptive patterns based on a sociotechnical perspective: it highlights how deceptive patterns can exploit and impact different aspects related to information and communication platforms (from hardware to language) and human information functions (from comprehension capacity to expectations and beliefs) in an interconnected manner. This work also shows how the academic community tends to focus more on aspects of human functions at the expense of the technical platform, which can result in incomplete analyses or conclusions that overlook important points, either social or technical, due to the systemic nature of deceptive patterns.

It is also hoped that the catalog will help users become aware of the design patterns created to deceive and manipulate them, thereby reducing the negative impacts on their lives. It can also raise awareness among interface design professionals about the real-world implications of their work and the importance of ethical practices. After the construction of the catalog and the preliminary evaluation of its ease of use and perceived usefulness, future work will involve assessing whether the catalog can support user awareness through a case study with users. Expanding the website to include more self-explanatory deceptive patterns and a quiz with questions and answers will possibly engage users further and reinforce the content learned from the catalog.

This work provided a broad overview of deceptive patterns, examining the works being done in the field. As future work, it is possible to use the Semiotic Framework to analyze this topic in greater depth, whether by analyzing other specific deceptive patterns or delving deeper into each layer to understand the internal mechanisms of deceptive patterns better. Alternatively, a broader literature review focused on each layer could be conducted.

For future research, each layer could be explored in depth to reveal more details about the mechanics of deceptive patterns. A deeper look into Human Information Functions (Social World, Pragmatic, Semantic) could help investigate the effects of the technical aspects of deceptive patterns in different contexts and among different audiences, considering how socioeconomic and cultural characteristics relate to deceptive patterns.

A deeper analysis of the Technical Platform (syntactic, empirical, physical world), in turn, could help understand how deceptive patterns are constructed and disseminated and aid in discovering common characteristics and structures that could be used to identify deceptive patterns manually or automatically. A better understanding of the phenomenon enables socially responsible ways to mitigate its practices and effects.

It will also be possible to conduct a more situated technical analysis of systems and websites of Brazilian companies, identifying which use unethical strategies and which do not, and perhaps create a certification seal for those that strive to create interfaces free of deceptive patterns.

\section*{Declarations}



Grammarly AI-review tool was used to help improve the English language. No other AI-generated content is used.


\begin{funding}
This research was partially funded by CAPES and the Academic Excellence Program (PROEX). Roberto Pereira holds a CNPq Research Productivity Scholarship (\#306423/2023-5). The authors especially thank their colleagues from the IHC Research Lab at UFPR.
\end{funding}

\begin{contributions}
According to the CRediT Taxonomy (\href{https://credit.niso.org/}{https://credit.niso.org/}) all authors contributed to the Conceptualization and Formal Analysis of this study. LB was responsible for the original draft's Investigation, Visualization, and Writting. RP was responsible for the Supervision, Validation, and Review \& Editing. AP also helped on Review \& Editing. All authors read and approved the final manuscript.

\end{contributions}



\begin{cds}
We acknowledge the importance of diversity in academic citations and have tried to include work from scholars of different backgrounds. Our citation practices aim to reflect contributions from researchers of various gender identities, racial and ethnic groups, and geographical locations. However, we recognize the limitations in the demographic data available for all cited authors and encourage further research on equitable citation practices.


\end{cds}

\bibliographystyle{apalike-sol}
\balance
\bibliography{refs}

\end{document}